\documentclass[prd,twocolumn,superscriptaddress,altaffilletter,amssymb,showpacs,nofootinbib]{revtex4}
\usepackage[dvips]{graphicx}
\usepackage{amsmath}
\usepackage{graphicx}
\usepackage{amsmath,amssymb}        
\usepackage{epsfig}
\usepackage{bm}
\usepackage{amsfonts}
\usepackage{amsmath,amssymb}
\usepackage{dcolumn}
\PassOptionsToPackage{linktocpage}{hyperref}
\usepackage[draft=false]{hyperref}
\usepackage{threeparttable}
\usepackage[draft=false]{hyperref}
\usepackage{threeparttable}
\newcommand{\sfrac}[2]{{\textstyle{#1\over#2}}}
\def\case#1/#2{\textstyle\frac{#1}{#2}}

\newcommand{\be}{\begin{equation}}
\newcommand{\ee}{\end{equation}}
\newcommand{\ben}{\begin{eqnarray}}
\newcommand{\een}{\end{eqnarray}}

\def\e{\mathbf{e}}
\def\sp{\sigma_+}
\def\udot{\dot{u}}
\def\ex{e_1{}^1}
\def\ey{e_2{}^2}

\def\Ex{E_1{}^1}

\def\y{\vartheta}
\def\z{\varphi}

\begin{document}

\title{Dynamics of the anisotropic Kantowsky-Sachs geometries in $R^n$ gravity}

\author{Genly Leon}\email{genly@uclv.edu.cu}
\affiliation{Department of Mathematics,\\ Universidad Central de
Las Villas, Santa Clara CP 54830, Cuba}

\author{Emmanuel N. Saridakis}
\email{msaridak@phys.uoa.gr}
 \affiliation{College of Mathematics
and Physics,\\ Chongqing University of Posts and
Telecommunications Chongqing 400065, P.R. China }

\begin{abstract}
We construct general anisotropic cosmological scenarios governed
by an $f(R)$ gravitational sector. Focusing then on
Kantowski-Sachs geometries in the case of $R^n$-gravity, and
modelling the matter content as a perfect fluid, we perform a
detailed phase-space analysis. We find that at late times the
universe can result to a state of accelerating expansion, and
additionally, for a particular $n$-range ($2<n<3$) it exhibits
phantom behavior. Furthermore, isotropization has been achieved
independently of the initial anisotropy degree, showing in a
natural way why the observable universe is so homogeneous and
isotropic, without relying on a cosmic no-hair theorem. Moreover,
contracting solutions have also a large probability to be the
late-time states of the universe. Finally, we can also obtain the
realization of the cosmological bounce and turnaround, as well as
of cyclic cosmology. These features indicate that anisotropic
geometries in modified gravitational frameworks present radically
different cosmological behaviors comparing to the simple isotropic
scenarios.
\end{abstract}


\maketitle

\section{Introduction}

The observable universe is homogeneous and isotropic at a great
accuracy \cite{Komatsu:2010fb}, leading the large majority of
cosmological works to focus on homogeneous and isotropic
geometries. The explanation of these features, along with the
horizon problem, was the main reason of the inflation paradigm
construction \cite{inflation}. Although the last subject is well
explained, the homogeneity and the isotropy problems are, strictly
speaking, not fully solved, since usually one starts straightaway
with a homogeneous and isotropic Friedmann-Robertson-Walker (FRW)
metric, and examines the evolution of fluctuations. On the other
hand, the robust approach on the subject should be to start with
an arbitrary metric and show that inflation is indeed realized and
that the universe evolves towards an FRW geometry, in agreement
with observations. However, the complex structure of such an
approach allows only for a numerical elaboration \cite{Gol:89},
and therefore in order to extract analytical solutions many
authors start with one assumption more, that is investigating
anisotropic but homogeneous cosmologies. This class of geometries
was known since a long time ago \cite{Misner:1974qy}, and it can
exhibit very interesting cosmological behavior, either in
inflationary or in post-inflationary cosmology
\cite{Peebles:1994xt}. Finally, note that such geometries may be
relevant for the description of the black-hole interior
\cite{Conradi:1994yy}.

The most well-studied homogeneous but anisotropic geometries are
the Bianchi type (see \cite{Ellis:1968vb,Tsagas:2007yx} and
references therein) and the Kantowski-Sachs metrics
\cite{chernov64,KS66,Collins:1977fg}, either in conventional or in
higher-dimensional framework. Although some Bianchi models (for
instance the Bianchi IX one) are more realistic, their complicated
phase-space behavior led many authors to investigate the simpler
but very interesting Bianchi I, Bianchi III, and Kantowski-Sachs
geometries. In these types of metrics one can analytically examine
the rich behavior and incorporate also the matter content
\cite{Henneaux:1980ft,Leach:2006br,Louko:1987gg,Collins:1977fg,Byland:1998gx},
obtaining a very good picture of homogeneous but anisotropic
cosmology.

On the other hand the observable universe is now known to be
accelerating \cite{obs}, and this feature led physicists to follow
two directions in order to explain it. The first is to introduce
the concept of dark energy (see \cite{Copeland:2006wr} and
references therein) in the right-hand-side of the field equations,
which could either be the simple cosmological constant or various
new exotic ingredients \cite{quint,phant,quintom}. The second
direction is to modify the left-hand-side of the field equations,
that is to modify the gravitational theory itself, with the
extended gravitational theories known as $f(R)$-gravity  (see
\cite{Sotiriou:2008rp,DeFelice:2010aj} and references therein)
being the most examined case. Such an approach can still be in the
spirit of General Relativity since the only request is that the
Hilbert-Einstein action should be generalized (replacing the Ricci
scalar $R$ by functions of it) asking for a gravitational
interaction acting, in principle, in different ways at different
scales. Such extended theories can present very interesting
behavior and the corresponding cosmologies have been investigated
in detail
\cite{Woodard:2006nt,Nojiri:2003ft,AGPT,matterper1,Zhang05}.

In the present work we are interested in investigating homogeneous
but anisotropic cosmologies, focusing on Kantowski-Sachs type, in
a universe governed by $f(R)$-gravity. In particular, we perform a
phase-space and stability analysis of such a scenario, examining
in a systematic way the possible cosmological behaviors, focusing
on the late-time stable solutions. Such an approach allows us to
bypass the high non-linearities of the cosmological equations,
which prevent any complete analytical treatment, obtaining a
(qualitative) description of the global dynamics of these models.
Furthermore, in these asymptotic solutions we calculate various
observable quantities, such are the deceleration parameter, the
effective (total) equation-of-state parameter, and the various
density parameters. We stress that the results of anisotropic
$f(R)$ cosmology are expected to be different than the
corresponding ones of $f(R)$-gravity in isotropic geometries,
similarly to the differences between isotropic \cite{eddington}
and anisotropic \cite{harrison} considerations in General
Relativity. Additionally, the results are expected to be different
from anisotropic General Relativity, too. As we see, anisotropic
$f(R)$ cosmology can be consistent with observations.

The paper is organized as follows: In section \ref{sectionII} we
construct the cosmological scenario of anisotropic $f(R)$-gravity,
presenting the kinematical and dynamical variables,
particularizing on the $f(R) = R^n$ ansatz in the case of
Kantowski-Sachs geometry. Having extracted the cosmological
equations, in section \ref{sectionIII} we perform a systematic
phase-space and stability analysis of the system. Thus, in section
\ref{sectionIV} we analyze the physical implications of the
obtained results, and we discuss the cosmological behaviors of the
scenario at hand. Finally, our results are summarized in section
\ref{conclusions}.

\section{Anisotropic $f(R)$ cosmology}\label{sectionII}

In this section we present the basic features of anisotropic
Bianchi I, Bianchi III, and Kantowski-Sachs geometries. After
presenting the kinematical and dynamical variables in the first
two subsections, we focus on the Kantowski-Sachs geometrical
background and on the $R^n$-gravity in the last two subsections.

\subsection{The geometry and kinematical variables}\label{sectionII.1}

In order to investigate anisotropic cosmologies, it is usual to
assume an anisotropic metric of the form
\cite{Byland:1998gx,Coley:2008qd}: {\small{
\begin{equation}
 ds^2 = - N(t)^2 dt^2 + [\ex(t)]^{-2} dr^2
  + [\ey(t)]^{-2} [d\y^2 + S(\y)^2\,
   d\z^2],\label{metric}
\end{equation}}}
where $1/\ex(t)$ and $1/\ey(t)$ are the expansion scale factors.
The frame vectors in coordinate form are written as
\begin{eqnarray}
  &&  \e_0 = N^{-1} \partial_t
    ,\quad
    \e_1 = \ex \partial_r
        \nonumber\\
    &&    \e_2 = \ey \partial_\y
        ,\quad \
        \e_3 = \ey/S(\y) \partial_\z.
\end{eqnarray} \

The  metric \eqref{metric} can describe three geometric families,
that is
\begin{eqnarray}
S(\y ) &=&\left\{
\begin{array}{l}
\sin {\y }\hspace{0.7cm}{\rm for}\hspace{0.3cm}k=+1, \\
\y \hspace{1.2cm}{\rm for}\hspace{0.3cm}k=0, \\
\sinh {\y }\hspace{0.5cm}{\rm for}\hspace{0.3cm}k=-1,
\end{array}
\right. \
\end{eqnarray}
known respectively as Kantowski-Sachs, Bianchi I and Bianchi III
models.

In the following, we will focus on the Kantowski-Sachs geometry
\cite{KS66}, since it is the most popular anisotropic model, and
since all solutions are known analytically in the case of General
Relativity, even if some particular types of matter are coupled to
gravity \cite{Collins:1977fg,Byland:1998gx}. These are spatially
homogeneous spherically symmetric models
\cite{Goheer:2007wu,Coley:2008qd}, with 4 Killing vectors: $
    \partial_\z,\quad
    \cos \z \ \partial_\y - \sin \z \cot \y \ \partial_\z,\quad
        \sin \z \ \partial_\y + \cos \z \cot \y \ \partial_\z,\partial_x$ \cite{kramer}.

Let us now consider relativistic fluid dynamics (see for e.g.
\cite{carge73}) in such a geometry. For any given fluid 4-velocity
vector field $u^{\mu}$, the projection tensor\footnote{Covariant
spacetime indices are denoted by letters from the second half of
the Greek alphabet.}
$$h_{\mu \nu}=g_{\mu \nu}+u_\mu u_\nu$$ projects into the instantaneous
rest-space of a comoving observer. It is standard to decompose the
first covariant derivative $\nabla_\mu u_\nu$ into its irreducible
parts
$$\nabla_\mu u_\nu=-u_\mu\dot u_\nu+\sigma_{\mu \nu}+\frac{1}{3}\Theta h_{\mu \nu}-\omega_{\mu
\nu},$$ where $\sigma_{\mu \nu}$ is the symmetric and trace-free
shear tensor ($\sigma_{\mu \nu} = \sigma_{(\mu \nu)}$,
$\sigma_{\mu \nu}\,u^\nu = 0$, $\sigma^\mu{}_{\mu}= 0$),
$\omega_{\mu \nu}$ is the antisymmetric vorticity tensor
($\omega_{\mu \nu} = \omega_{[\mu \nu]}$, $\omega_{\mu \nu}\,u^\nu
= 0$) and $\udot_\mu$ is the acceleration vector defined as
$\udot_\mu=u^\nu\nabla_{\nu}u_{\mu}$ (a dot denotes the derivative
with respect to $t$). In the above expression we have introduced
also the volume expansion scalar. In particular,
\begin{equation}
\Theta= \nabla_\mu u^\mu,
\label{thetadef}
\end{equation}
 which defines a length scale $\ell$
along the flow lines, describing the volume expansion
(contraction) behavior of the congruence completely, via the
standard relation, namely
\begin{equation}
\Theta \equiv\sfrac{3\dot{\ell}}{\ell}.
\label{thetaelldef}
\end{equation} In cosmological
contexts it is customary to use the Hubble scalar $H=\Theta/3$
\cite{EU}. From these it becomes obvious that in FRW geometries,
$\ell$ coincides with the scale factor. Finally, it can be shown
that these kinematical fields are related through
\cite{carge73,EU}
\begin{align}
&\sigma_{\mu \nu}:=\dot u_{(\mu} u_{\nu)}+\nabla_{(\mu}
u_{\nu)}-\frac{1}{3} \Theta h_{\mu \nu}\\
& \omega_{\mu \nu}:= -u_{[\mu}\dot u_{\nu]}-\nabla_{[\mu}
u_{\nu]}.
\end{align}

Transiting to the synchronous temporal gauge, we can set $N$ to
any positive function of $t$, or simply $N=1$. Thus, we extract
the following restrictions on kinematical variables:
\begin{equation}
   \sigma^\mu_\nu = \text{diag}(0,-2\sp,\sp,\sp),\quad
    \omega_{\mu \nu} =0,\quad
    \end{equation}
     where
     \begin{equation}
     \label{sigmaplus}
      \sp=\frac13\frac{  d}{ {d} t}\left[\ln
      \frac{\ex}{\ey}\right].
       \end{equation}
Finally, note that the Hubble scalar can be expressed in terms of
$\ex$ and $\ey$ as
    \begin{equation}
    H=-\frac13\frac{  d}{ {d} t}\left[\ln \ex
(\ey)^2\right].
\end{equation}

\subsection{The action and dynamical variables}\label{sectionII.2}

Let us now construct $f(R)$ cosmology in such a geometrical
background. In the metric formalism the action for $f(R)$-gravity
is given by \cite{Sotiriou:2008rp,DeFelice:2010aj}
\begin{equation}
 {\cal S}_{\text{metric}}=\int  {d} x^4
\sqrt{-g}\left[f(R)-2\Lambda+{\cal L}^{(m)}\right]\label{action},
 \end{equation}
where $f(R)$ is a function of the Ricci scalar $R$, and  ${\cal
L}^{(m)}$ accounts for the matter content of the universe.
Additionally, we use the metric signature $(-1,1,1,1)$, 
Greek indices run from $0$ to $3$, and we impose the standard units in
which $c=8\pi G=1$. Finally, in the following, and without loss of
generality, we set the usual cosmological constant $\Lambda=0$.

The fourth-order equations obtained by varying the action
(\ref{action}) with respect to the metric write:
\begin{equation}
 G_{\mu \nu}=\frac{T_{\mu \nu}^{(m)}}{f'(R)}+T_{\mu \nu}^R,\label{EFE}
 \end{equation}
 where the prime
denotes differentiation with respect to $R$. In this expression
$T_{\mu \nu}^{(m)}$ denotes the matter energy-momentum tensor,
which is assumed to correspond to a perfect fluid with energy
density $\rho_m$ and pressure $p_m$. Additionally, $T_{\mu \nu}^R$
denotes a correction term describing a ``curvature-fluid''
energy-momentum tensor of geometric origin \cite{goswami,goheer}:
{
\begin{eqnarray}
T_{\mu \nu}^R=\frac{1}{f'(R)}\left[\frac{1}{2} g_{\mu
\nu}\left(f(R)-R f'(R)\right) \ \ \ \ \ \ \ \ \ \right.\nonumber\\
\left. \ \ \ \ \ \ \ \  \ \ \ \ \ \ + \nabla_\mu \nabla_\nu
f'(R)-g_{\mu \nu}\Box f'(R)\right], \label{feqs}
 \end{eqnarray}}
where $\nabla_\mu$ is the covariant derivative associated to the
Levi-Civita connection of the metric and $\Box\equiv
\nabla^\mu\nabla_\mu.$ Note that in the last two terms of the
right hand side there appear fourth-order metric-derivatives,
justifying the name ``fourth order gravity'' used for this class
of theories \cite{DeFelice:2010aj}. By taking the trace of
equation (\ref{EFE}) and re-ordering terms one obtains the ``trace
equation'' (equation (5) in section IIA of
\cite{Capozziello:2009nq})
\begin{equation}
3 \Box f'(R)+R f'(R)-2 f(R)=T\label{Trace},
\end{equation} where $T=T_\mu^\mu$ is the trace of the energy-momentum tensor of ordinary matter.

In the phenomenological fluid description of a general matter
source, the standard decomposition of the energy-momentum tensor
$T_{\mu \nu}$ with respect to a timelike vector field $u^\mu$ is
given by
\begin{equation}
T_{\mu \nu}=\mu u_\mu u_\nu +2 q_{(\mu} u_{\nu)} +P h_{\mu
\nu}+\pi_{\mu \nu},
\end{equation}
 where $\mu$ denotes the energy density scalar, $P$ is the isotropic pressure
 scalar,
$q_{\mu}$ is the energy current density vector ($q_\mu\,u^\mu =
0$) and $\pi_{\mu\nu}$ is the trace-free anisotropic pressure
tensor ($\pi_{\mu \nu}u^\nu=0,\, \pi_\mu^\mu=0, \pi_{\mu
\nu}=\pi_{\nu \mu}$).

The matter fields need to be related through an appropriate
thermodynamical equation of state in order to provide a coherent
picture of the physics underlying the fluid spacetime scenario.
Applying this covariant decomposition to the ``curvature-fluid''
energy-momentum tensor \eqref{feqs} we obtain
\begin{align}
&\mu=-\frac{1}{2}\left[\frac{f(R)-R
f'(R)+6 H\frac{ {d}}{ {d}t}f'(R)}{f'(R)}\right]\nonumber\\
& P=-\frac{1}{2}\left[\frac{-f(R)+R f'(R)-4 H\frac{ {d}}{
{d}t}f'(R)-2\frac{ {d}^2}{
{d}t^2}f'(R)}{f'(R)}\right].\label{matterfields1}
\end{align}
Finally, the anisotropic pressure tensor is given by $\pi^\mu_\nu
= \text{diag}(0,-2\pi_+,\pi_+,\pi_+)$, where
\begin{equation}
\pi_+=-\frac{\frac{ {d}}{
{d}t}f'(R)}{f'(R)}\sp.\label{matterfields2}
\end{equation}

\subsection{The cosmological equations}\label{sectionII.3}

Let us now present the cosmological equations in the homogeneous
and anisotropic Kantowski-Sachs metric. In particular, the
Einstein's equations  \eqref{EFE} in the Kantowski-Sachs metric
 read
\begin{align}
& 3\sp^2-3 H^2-{}^2\!K =-\mu-\frac{\rho_m
}{f'(R)},\nonumber\\
 &-3(\sp+H)^{2}-2\dot\sp-2\dot H
-{}^2\!K=\frac{p_m}{f'(R)}+P-2\,\pi_+,\nonumber\\
&-3\sp^{2}+3\sp H -3H ^{ 2}+\dot\sp -2\dot H=
\frac{p_m}{f'(R)}+P+\pi_+.\label{varyingaction}
\end{align}
In these expressions $\rho_m$ and $p_m$ are the energy density and
pressure of the matter perfect fluid, and their ratio gives the
matter equation-of-state parameter
\begin{equation}
w=\frac{p_m}{\rho_m}.
\end{equation}
Furthermore, ${}^2\!K$ is the Gauss curvature of the 3-spheres
\cite{EU}
  given by
 \begin{equation}
    {}^2\!K = (\ey)^2,
\end{equation}
and its evolution equation writes
\begin{equation}
 \dot {{}^2\!K}=-2 (\sp +H) \left(
{{}^2\!K}\right)\label{propK}.
 \end{equation}
 Additionally, the evolution equation for $\ex$ reads (see equation (42) in section 4.1 of \cite{Coley:2008qd})
 \begin{equation} \dot{e}_1{}^1 =-
\left(H-2\sp\right)\ex\label{aux}.
\end{equation}
Finally, observing the form of the first of equations
(\ref{varyingaction}) one can define the various density
parameters of the scenario at hand, namely \cite{REZA} the
curvature one
\begin{equation}
\label{Omegas1} \Omega_k\equiv -\frac{{{}^2\!K}}{3H^2},
\end{equation}
the matter one
\begin{equation}
\label{Omegas2}
 \Omega_{m}\equiv\frac{\rho_m}{3 H^2
f'(R)},
\end{equation}
the ``curvature-fluid'' one
\begin{equation}
\label{Omegas3}
 \Omega_{\text{curv.fl}}\equiv\frac{\mu}{3
H^2},
\end{equation}
and the shear one
\begin{equation}
\Omega_\sigma\equiv\left(\frac{\sp}{H}\right)^2\label{Omegas4},
\end{equation}
 satisfying $\Omega_k+\Omega_m+\Omega_{\text{curv.fl}}+\Omega_\sigma=1$.

Now, the trace equation \eqref{Trace} in the Kantowski-Sachs
metric reads
\begin{equation}
 -3\frac{{d}^2}{{d}t^2}f'(R)-9 H \frac{{d}}{{d}t}f'(R)+R f'(R)-2 f(R)
 = -\rho_m+3 p_m,\label{Tr}
 \end{equation}
and the Ricci scalar writes
\begin{equation}
\label{Riccidef}
  R= 12H ^{2}+6
\sp^{2}+6\dot H +2{}^2\!K.
  \end{equation}

At this stage we can reduce equations \eqref{varyingaction} along
with \eqref{Tr},  with respect to $\dot H,\,\dot \sp$ and $
{}^2\!K$  \cite{Rippl96},  acquiring the Raychaudhuri  equation
{\small{
\begin{eqnarray}
 \dot H=-H^2-2\sp^2-\frac{1}{6
f'(R)}[\rho_m+3p_m]-\frac{H}{2f'(R)}
\frac{{d}}{{d}t}f'(R)\nonumber\\
 +
\frac{1}{6}\left[R-\frac{f(R)}{f'(R)}
-\frac{3}{f'(R)}\frac{{d}^2}{{d}t^2}f'(R)\right], \ \label{Raych}
\end{eqnarray}}}
the shear evolution
\begin{eqnarray}
 \dot\sp=-\sp^2-3H\sp +H^2-\frac{\rho_m}{3f'(R)} \ \ \ \ \ \   \ \ \ \ \ \ \nonumber\\
 \ \ \ \ \ \ \ -\frac{1}{6}\left[R-\frac{f(R)}{f'(R)}\right]
+\frac{(H-\sp)}{f'(R)}\frac{{d}}{{d}t}f'(R),\label{shear}
\end{eqnarray}
and the Gauss constraint {\small{
\begin{equation}
\label{gauss111}
 {}^2\!K=3\sp^2-3H^2+\frac{\rho_m}{f'(R)}
+\frac{1}{2}\left[R-\frac{f(R)}{f'(R)}\right]-\frac{3H}{f'(R)}\frac{
{d}}{ {d}t}f'(R).
\end{equation}}}
It proves convenient to use the trace equation \eqref{Tr} to
eliminate the derivative $\frac{ {d}^2}{ {d}t^2}f'(R)$ in
\eqref{Raych},  obtaining a simpler form of Raychaudhuri equation,
namely
\begin{equation}
\dot H=-H^2-2\sp^2-\frac{\rho_m}{3 f'(R)}+\frac{f(R)}{6
f'(R)}+\frac{1}{f'(R)}H \frac{ {d}}{ {d}t}f'(R)\label{Raych2}.
\end{equation}
Furthermore, the Gauss constraint can alternatively be expressed
as
\begin{eqnarray}
\left[H+\frac{1}{2}\frac{\frac{ {d}f'(R)}{ {d}t}}
{f'(R)}\right]^2+\frac{1}{3}{}^2\!K=\sp^2+\frac{\rho_m}{3f'(R)}
\ \ \ \ \   \ \ \ \ \ \nonumber\\
\ \ \ \ \   \ \ \ \ \  \ \ \
+\frac{1}{6}\left[R-\frac{f(R)}{f'(R)}\right]+\frac{1}{4}\left[\frac{\frac{
{d}f'(R)}{ {d}t}} {f'(R)}\right]^2.\label{Gauss}
\end{eqnarray}

In summary, the cosmological equations of $f(R)$-gravity in the
Kantowski-Sachs background are the ``Raychaudhuri equation''
\eqref{Raych2}, the shear evolution \eqref{shear}, the trace
equation \eqref{Tr}, the Gauss constraint \eqref{Gauss}, the
evolution equation for the 2-curvature ${}^2\!K$ \eqref{propK},
and the evolution equation for $\ex$ \eqref{aux}. Finally, these
equations should be completed by considering the evolution
equations for matter sources.

\subsection{$R^n$-gravity}\label{sectionII.4}

In this subsection we specify the above general cosmological
system. In particular, in order to continue we have to make an
assumption concerning the function $f(R)$. We impose an ansatz of
the form $f(R)=R^n$
\cite{Sotiriou:2008rp,DeFelice:2010aj,Carloni05,Goheer:2007wu},
since such an ansatz does not alter the characteristic length
scale (and General Relativity is recovered when $n=1$), and it
leads to simple exact solutions which allow for comparison with
observations \cite{Capozzi02,Capozzi03}. Additionally, following
\cite{goswami,goheer} we consider that the parameter $n$ is
related to the matter equation-of-state parameter through
\begin{equation}
\label{wn}
 n=\frac{3}{2}(1+w).
\end{equation}    Such a choice is imposed by the requirement
of the existence of an Einstein static universe in FRW
backgrounds, which leads to a severe constraint in $f(R)$, namely
$f(R)=2\Lambda+R^{\frac{3}{2}(1+w)}$  with $ w\neq -1$. Inversely,
it can be seen as the equation-of-state parameter of ordinary
matter is a fixed function of $n$, if we desire Einstein static
solutions to exist. The existence of such a static solution (in
practice as a saddle-unstable one) is of great importance in every
cosmological theory, since it connects the Friedmann
matter-dominated phase with the late-time accelerating phase, as
required by observations \cite{goswami,goheer}. Thus, if we relax
the condition $n=\frac{3}{2}(1+w)$ in $R^n$-gravity, in general we
cannot obtain the epoch-sequence of the universe. Finally, note
that the constraint $-1/3\leq w\leq1$ that arises from the
satisfaction of all the energy conditions for standard matter,
imposes bounds on $n$, namely
 $n\in[1,3]$, with the most interesting cases being those of
 dust ($w=0$, $n=3/2$) and radiation fluid
($w=1/3$, $n=2$).

The cosmological equations for $R^n$-gravity, in the homogeneous
and anisotropic Kantowski-Sachs metric (\ref{metric}), are
obtained by specializing equations \eqref{Raych2}, \eqref{shear},
\eqref{Gauss}, \eqref{Tr}, \eqref{propK}, \eqref{aux} and consider
the conservation equation for standard matter.

Equations \eqref{Raych2}, \eqref{shear}, \eqref{Gauss} and
\eqref{Tr}, become respectively
 \begin{eqnarray}
 \label{cosmeq1}
  \dot H+ H^2=
-2\sp^2-\frac{\rho_m}{3 n R^{n-1}}+\frac{1}{6n} R+(n-1)
H\frac{\dot R}{R},
 \end{eqnarray}
  \begin{eqnarray} \label{cosmeq2}
\dot\sp=-\sp^2-3 H \sp +H^2 -\frac{\rho_m}{3 n
R^{n-1}}-\frac{n-1}{6 n}
R\ \nonumber\\
-(n-1)\left(\sp-H\right)\frac{\dot R}{R},
 \end{eqnarray}
   \begin{eqnarray}
 \left(H+\frac{n-1}{2}\frac{\dot R}{R}\right)^2+\frac{1}{3}
{}^2\!K=\sp^2+\frac{\rho_m}{3 n R^{n-1}}\nonumber\\+
\frac{n-1}{6n} R+\frac{(n-1)^2}{4}\frac{\dot
R^2}{R^2},\label{Gauss2}
\end{eqnarray}
   \begin{equation}
 \frac{\ddot
R}{R}=\frac{n-2}{3 n(n-1)} R+\frac{2(2-n)}{3
n(n-1)}\frac{\rho_m}{R^{n-1}}-3 H\frac{\dot R}{R}-(n-2)\frac{\dot
R^2}{R^2}.
 \label{cosmeq4}
 \end{equation}
Additionally, we consider the evolution equation for the
2-curvature ${}^2\!K$ given by \eqref{propK}, as well as the
matter conservation equation
\begin{equation}
 \label{cosmeq5}
 \dot \rho_m=-2n H \rho_m.
\end{equation}
Finally, in order to close the equation-system we consider also
the propagation equation \eqref{aux}.

\section{Phase-space analysis}\label{sectionIII}

In the previous section we formulated $f(R)$-gravity in the case
of the homogeneous and anisotropic Kantowski-Sachs geometry,
focusing on the $f(R)=R^n$ ansatz. Having extracted the
cosmological equations we can investigate the possible
cosmological behaviors and discuss the corresponding physical
implications by performing a phase-space analysis. Such a
procedure bypasses the complexity of the cosmological equations
and provides us the understanding of the dynamics of these
scenarios.

\subsection{The dynamical system}\label{sectionIII.1}

In order to perform the phase-space and stability analysis of the
models at hand, we have to transform the cosmological equations
into an autonomous dynamical system \cite{Copeland:1997et}.
However, since the present system is more complicated, in order to
avoid ambiguities related to the non-compactness at infinity we
define compact variables that can describe both expanding and
collapsing models \cite{Campos01a,Goheer:2007wu}. This will be
achieved by introducing the auxiliary variables
\cite{Solomons:2001ef,Goheer:2007wx}:
\begin{eqnarray}
 &&Q=\frac{H}{D},\ \ \ \
\Sigma=\frac{\sp}{D},\nonumber\\
 &&x=\frac{(n-1) \dot R}{2 R D},\ \ \ \
y=\frac{(n-1)R}{6 n D^2},\ \ \ \
 z=\frac{\rho_m}{3 n R^{n-1}D^2},\nonumber\\
&&K=\frac{{}^2\!K}{3 D^2},\ \ \ \ \Ex=\frac{\ex}{D},
\label{auxiliary}
\end{eqnarray}
where we have defined
 \begin{equation}
 \label{Ddef}
D=\sqrt{\left(H+\frac{n-1}{2}\frac{\dot R}{R}\right)^2+\frac{1}{3}
{}^2\!K}.
 \end{equation}
Furthermore, we introduce the time variable $\tau$ through
\begin{equation}
 {d}\tau=\left(\frac{D}{n-1}\right) {d} t,
\end{equation}
and from now on primes will denote derivatives with respect to
$\tau$. Finally, note that in the following we focus on the
general case $n\neq1$. The dynamical investigation of
Kantowsky-Sachs model in General Relativity (that is for $n=1$)
has been performed in \cite{Coley:2003mj}.

In terms of these auxiliary variables the Gauss constraint
(\ref{Gauss2}) becomes
\begin{equation}
\label{constr1}
 x^2+y+z+\Sigma^2=1.
 \end{equation}
Moreover,  the $D$-definition (\ref{Ddef}) becomes
\begin{equation}
\label{constr2}
 \left(Q+x\right)^2+K=1.
\end{equation}
Finally, from the definitions (\ref{auxiliary}) we obtain the
bounds $y\geq 0,\, z\geq 0$ and $K\geq 0$. Therefore, we conclude
that the auxiliary variables must be compact and lie at the
intervals
 \begin{eqnarray}
&& Q\in[-2,2],\ \ \ \ \Sigma\in[-1,1],\nonumber\\
 &&  x\in [-1,1],\ \ \ \
y\in[0,1],\ \ \ \ z\in[0,1]\nonumber\\
&& K\in[0,1],
\end{eqnarray}
while $\Ex$ is un-constrained.

In summary, using the dimensionless auxiliary variables
(\ref{auxiliary}), along with the two constraints (\ref{constr1})
and (\ref{constr2}), we reduce the complete cosmological system to
a five-dimensional one given by:
\begin{widetext}
\begin{eqnarray}
  && Q'= (1 -n  ) \Sigma Q^3-(n-1) \left[-3 x^2+2
\Sigma x+(n-3) \Sigma ^2+n \left(x^2+y-1\right)+1\right]
Q^2\nonumber\\
&&\ \ \ \ \ \ \ -(n-1) \left\{x
   \left[-3 x^2+(x-3 \Sigma ) \Sigma +n \left(x^2+\Sigma ^2+y-1\right)-1\right]-\Sigma
   \right\}
   Q\nonumber\\
\ &&\ \ \ \ \ \ \
   -x^2+\Sigma ^2+n
   \left(x^2-\Sigma ^2+y-1\right)+1, \label{eqQ}
   \end{eqnarray}
\begin{eqnarray}
&&\Sigma'=-(n-3) (n-1) (Q+x) \Sigma ^3-(n-1) (Q+x-1) (Q+x+1)
\Sigma
^2\nonumber\\
&&\ \ \ \ \ \ \ -(n-1) (Q+x) \left[(n-3) \left(x^2-1\right)+n
y\right] \Sigma +(n-1)
   \left[(Q+x)^2-1\right],\label{eqSigma}
       \end{eqnarray}
    \begin{eqnarray}
&&x'=-(n-3) (n-1) x^4-(n-1) \left[(n-3) Q+\Sigma \right] x^3-(n-1)
\left[-3 \Sigma ^2+2 Q \Sigma +n \left(\Sigma
^2+y-2\right)+5\right]
   x^2\nonumber\\
 && \ \ \ \ \ \ \
   -(n-1) \left\{Q \left[(Q-3 \Sigma ) \Sigma +n \left(\Sigma ^2+y-1\right)+3\right]-\Sigma
   \right\}
   x+(n-2) \left[-\Sigma ^2+n
   \left(\Sigma ^2+y-1\right)+1\right],\label{eqx}
       \end{eqnarray}
    \begin{eqnarray}
& y'=-2 (n-1) n (Q+x)
   y^2+y \left\{-2 \left[(n-4) n+3\right] (Q+x) \Sigma ^2-2 (n-1) (Q+x-1)
(Q+x+1) \Sigma \right. \nonumber \\ & \left.  -2 \left\{(Q+x)
\left(x^2-1\right) n^2+\left(-4 x^3-4 Q
   x^2+2 x+Q\right) n+x [3 x
   (Q+x)-2]\right\}\right\},\label{eqy}
    \end{eqnarray}
    \begin{eqnarray}
& \left(\Ex\right) '=\Ex (n-1) \left\{-(n-3) (Q+x) \Sigma
^2-\left[(Q+x)^2-3\right] \Sigma +(-Q-x) \left[-3 x^2+n
   \left(x^2+y-1\right)+1\right]\right\}
 \label{Aux}.
  \end{eqnarray}
\end{widetext}

Proceeding forward, as we observe, equation \eqref{Aux} decouples
from \eqref{eqQ}-\eqref{eqy}.  Thus, if one is only interested in
the dynamics of Kantowsky-Sachs $f(R)$-models, all the relevant
information of the system is given by the remaining equations.
Therefore, we can reduce our analysis to the system
(\ref{eqQ})-(\ref{eqy}), defined in the compact phase-pace
\begin{widetext}
\begin{equation}
 \Psi=\left\{x^2+y+\Sigma^2\leq 1, |Q+x|\leq 1,
x\in [-1,1],\, y\in[0,1],\,\Sigma\in[-1,1], Q\in[-2,2]\right\}.
\end{equation}
\end{widetext}

Finally, while for the purpose of the present work it is adequate
to investigate the system \eqref{eqQ}-\eqref{eqy}, leaving outside
the decoupled equation \eqref{Aux}, for completeness we study
equation \eqref{Aux} in Appendix \ref{subsE1}.

\subsection{Invariant sets and critical points}\label{sectionIII.2}

Using the auxiliary variables (\ref{auxiliary}) the cosmological
equations of motion (\ref{cosmeq1})-(\ref{cosmeq5}), were
transformed into the autonomous form (\ref{eqQ})-(\ref{Aux}), that
is in a form
 $\textbf{X}'=\textbf{f(X)}, $ where $\textbf{X}$ is the column
vector constituted by the auxiliary variables and $\textbf{f(X)}$
the corresponding  column vector of the autonomous equations. The
critical points $\bf{X_c}$ are extracted satisfying $\bf{X}'=0$,
and in order to determine the stability properties of these
critical points we expand around $\bf{X_c}$, setting
$\bf{X}=\bf{X_c}+\bf{U}$ with $\textbf{U}$ the perturbations of
the variables considered as a column vector. Thus, up to first
order we acquire $ \textbf{U}'={\bf{\Xi}}\cdot \textbf{U}, $ where
the matrix ${\bf {\Xi}}$ contains the coefficients of the
perturbation equations. Therefore, for each critical point, the
eigenvalues of ${\bf {\Xi}}$ determine its type and stability. In
particular, eigenvalues with negative (positive) real parts
correspond to a stable (unstable) point, while eigenvalues with
real parts of different sign correspond to a saddle point. Lastly,
when at least one eigenvalue has zero real part, the corresponding
point is a non-hyperbolic one.

Now, there are several invariant sets, that is areas of the
phase-space that evolves to themselves under the dynamics, for the
dynamical system \eqref{eqQ}-\eqref{eqy}. There are two invariant
sets given by $Q+x=\pm1$, corresponding to $K=0$, that is
$\ey/D=0.$ However, the solutions of the evolution equations
inside this invariant set do not correspond to exact solutions of
the field equations, since the frame variables has to satisfy
$\det (\e_a)\neq 0$. Nevertheless, it appears that this invariant
set plays a fundamental role in describing the asymptotic behavior
of cosmological models.

Another invariant set is $y=0$, that is $\rho_m=0,$ $R\equiv 12
H^2+6 \dot H+6 \sigma^2+2 {}^2\!K=0$, provided $D$ is finite, or
$\rho_m=0,$ $R=R_0$, with $R_0$ a constant, provided $D\rightarrow
\infty$. This invariant set contains vacuum (Minkowsky) and static
models.

Another invariant set appears in the case of radiation background
($n=2, w=1/3$), namely that of $x=0.$

Finally, we have identified the invariant set $x^2+y+\Sigma^2=1$,
which contains cosmological solutions without standard matter.

\subsection{Local analysis of the dynamical system}
\label{localanal}

The system (\ref{eqQ})-(\ref{eqy}) admits two circles of critical
points, given by $C_\epsilon: \Sigma^2+Q^2-2 \epsilon Q=0,\,
x+Q=\epsilon,\,y=0$ (we use the notation $\epsilon=\pm1$). They
exist for $n\in [1,3]$, they are located in the boundary of
$\Psi$, and they correspond to solutions in the full phase space,
satisfying $K=y=z=0$. In order to be more transparent, let us
consider the parametrization
\begin{equation}
C_\epsilon:=\left\{\begin{array}{c}
  Q=\epsilon+\sin u \\
  \Sigma=\cos u \\
  x=-\sin u. \\
\end{array},\, u\in [0,2\pi]\right.
\end{equation}

For all the critical points we calculate the eigenvalues of the
perturbation matrix $\bf{\Xi}$, which will determine their
stability. These eigenvalues, evaluated at $C_\epsilon$, are
\begin{eqnarray}
 &&\{
 0,\, -2(n-1)(\cos
u-2\epsilon),\nonumber\\
&&\ \ -2(n-1)\left[(n-2)\sin u-\epsilon (3-n)\right],\nonumber\\
&&\ \ 2(n-2)\sin u+6\epsilon (n-1) \}.
\end{eqnarray}
 For the values of $u$ such
that there exists only one zero eigenvalue, the curves are
actually ``normally hyperbolic''\footnote{ Since we are dealing
with curves of critical points, every such point has necessarily
at least one eigenvalue with zero real part. ``Normally
hyperbolic'' means that the only eigenvalues with zero real parts
are those whose corresponding eigenvectors are tangent to the
curve \cite{normally}.}, and thus we can analyze the stability by
analyzing the sign of the real parts of the non-null eigenvalues
\cite{normally}. Therefore, we deduce that:
\begin{itemize}
\item No part of $C_+$ ($C_-$) is a stable (unstable). Thus, any
part of $C_+$ ($C_-$) is either unstable (stable) or saddle.
 \item
For $-\frac{1}{3}<w\leq -\frac{1}{6},$ one branch of $C_-$
($\frac{3(3w+1)}{3w-1}<\sin (u)\leq 1$) is stable, and one branch
of $C_+$ ($\frac{3(3w+1)}{3w-1}<\sin (-u)\leq 1$) is unstable.
\item For $-\frac{1}{6}<w<\frac{2}{3},$ the whole $C_-$ is stable
and the whole $C_+$ is unstable.
 \item For $\frac{2}{3}\leq w\leq 1,$ one
branch of $C_-$ ($\frac{3(w-1)}{3 w-1}<\sin (u)\leq 1$) is stable,
and one branch of $C_+$ ($\frac{3(w-1)}{3 w-1}<\sin (-u)\leq 1$)
is unstable.
\end{itemize}

Note that amongst the curves $C_\epsilon$, we can select the
representative critical points labelled by ${\cal
N}_\epsilon:=\left(Q=0,\,\Sigma=0,\, x=\epsilon,\, y=0\right)$ and
$\mathcal{L}_\epsilon:=\left(Q=2\epsilon,\,\Sigma=0,
x=-\epsilon,\,y=0\right)$ described in \cite{goheer}. In our
notation: ${\cal L}_-=C_-|_{u=3\pi/2}, {\cal N}_+=C_+|_{u=3\pi/2}$
and ${\cal N}_-=C_-|_{u=\pi/2}, {\cal L}_+=C_+|_{u=\pi/2}.$
Moreover, and contrary to the investigation of \cite{goheer}, we
obtain four new critical points, which are a pure result of the
anisotropy. They are labelled by
$P_1^\epsilon:=\left(Q=\epsilon,\,\Sigma=\epsilon,\, x=0,\,
y=0\right)$ and
$P_2^\epsilon:=\left(Q=\epsilon,\,\Sigma=-\epsilon,\, x=0,\,
y=0\right)$, and obviously $P_1^+=C_+|_{u=0}, P_1^-=C_-|_{u=\pi}$
and $P_2^+=C_+|_{u=\pi}, P_2^-=C_-|_{u=0}.$ Thus, it is easy to
see that:
\begin{itemize}
 \item For $-\frac{1}{6}<w<\frac{2}{3},$ ${\cal L}_+$
is unstable and ${\cal L_-}$ is stable.
 The critical point ${\cal
N}_+$ (${\cal N}_-$) is always unstable (stable) except in the
case $w= -\frac{1}{3}.$ These results match with the results
obtained in \cite{goheer}.
 \item The critical points
$P_1^\epsilon$ and $P_2^\epsilon$ exist always. They are
non-hyperbolic, presenting an 1-dimensional center manifold
tangent to the line $x+Q=0$. $P_{1,2}^-$ ($P_{1,2}^+$) have a 3D
stable (unstable) manifold provided $-\frac{1}{3}< w<1.$
\end{itemize}
The curves of critical points $C_\epsilon$, and the representative
critical points ${\cal N}_\epsilon$, $\mathcal{L}_\epsilon$,
$P_1^\epsilon$ and $P_2^\epsilon$, are enumerated in Table
\ref{tab1b}.

\begin{table*}
    \centering
\begin{tabular}{ccccccc}
  \hline   \hline
  Cr./Curve & $Q$ & $\Sigma$ & $x$ & $y$ & Existence& Stability\\
  \hline \hline
  $\mathcal{N_+}$ &  0  & 0  & 1
  & 0 & always & unstable
 \\ \hline
  $\mathcal{N_-}$ &  0  &  0 & -1
  & 0 & always & stable
 \\ \hline
    $\mathcal{L_+}$ &  2  &  0 & -1
  & 0 & always & unstable  for $1.25\leq n\leq2.5$
 \\ \hline
  $\mathcal{L_-}$ &  -2  &  0 & 1
  & 0 & always & stable  for $1.25\leq n\leq2.5$
 \\ \hline
      $P_1^+$ & $1$ & $1$ & $0$ & $0$ & $1< n\leq 3$ &{\text{non-hyperbolic with $3D$ unstable manifold}}\\ \hline
  $P_1^-$ & $-1$ & $-1$ & $0$ & $0$ & $1< n\leq 3$ &{\text{non-hyperbolic with $3D$ stable manifold\ \ \ }}\\ \hline
  $P_2^+$ & $1$ & $-1$ & $0$ & $0$ & $1< n\leq 3$ &{\text{non-hyperbolic with $3D$ unstable manifold}} \\ \hline
   $P_2^-$ & $-1$ & $1$ & $0$ & $0$ & $1< n\leq 3$ &{\text{non-hyperbolic with $3D$ stable manifold\ \ \  }}\\ \hline
  ${\cal C}_+$ & $1+\sin u$ &  $\cos u$ & $-\sin u$ & 0  & always & unstable for $1.25\leq n\leq2.5$  \\ \hline
    ${\cal C}_-$ & $-1+\sin u$ &  $\cos u$ & $-\sin u$ & 0  & always & stable for $1.25\leq n\leq2.5$\\ \hline
  \hline
\end{tabular}
\caption{{\label{tab1b}}The curves of critical points
$C_\epsilon$, and the representative critical points ${\cal
N}_\epsilon,$  $\mathcal{L}_\epsilon$, $P_1^\epsilon$ and
$P_2^\epsilon$ of the cosmological system (we use
$\epsilon=\pm1$). $u$ varies in $[0,2\pi].$ For more details on
the stability of $C_\epsilon$ see the text.}
\end{table*}

Until now we have extracted and analyzed the stability of the
curves of critical points $C_\epsilon$ of the system
(\ref{eqQ})-(\ref{eqy}), and their representative critical points.
However, the system (\ref{eqQ})-(\ref{eqy}) admits additionally
ten isolated critical points enumerated in Table \ref{tab1}, where
we also present the necessary conditions for their existence.
\footnote{Strictly speaking, the system (\ref{eqQ})-(\ref{eqy})
admits two more critical  points with coordinates $Q=\pm2,
\Sigma=\pm 1, x=0, y=0.$ However, they are unphysical since they
satisfy $|Q+x|> 1$ and thus ${}^2\!K\equiv \left(\ey\right)^2<0.$}


\begin{table*}
    \centering
\begin{tabular}{ccccccc}
  \hline   \hline
  Cr. P. & $Q$ & $\Sigma$ & $x$ & $y$ & Existence& Stability\\
  \hline \hline
  $\mathcal{A}_+$ & $\frac{2n-1}{3(n-1)}$ & $0$ & $\frac{n-2}{3(n-1)}$
  & $\frac{(2n-1)(4n-5)}{9(n-1)^2}$ & $1.25 \leq n \leq 3$ & \text{stable for} $n_+< n < 3$
\\\vspace{0.2cm}
     &   &  &
  &  &  & \text{saddle for} $1.25\leq n\leq n_+$
 \\ \hline
  $\mathcal{A}_-$ & ${-}\frac{2n-1}{3(n-1)}$ & $0$ & $-\frac{n-2}{3(n-1)}$
  & $\frac{(2n-1)(4n-5)}{9(n-1)^2}$ & $1.25 \leq n \leq 3$ & \text{unstable for} $n_+< n < 3$ \\\vspace{0.2cm}
      &   &  &
  &  &  & \text{saddle for} $1.25\leq n\leq n_+$
 \\ \hline
  ${\cal B}_+$ & $\frac{1}{3-n}$ & $0$ & $\frac{n-2}{n-3}$  & $0$ & $1< n\leq 2.5$&{\text{saddle}} \\ \hline
    ${\cal B}_-$ & $-\frac{1}{3-n}$ & $0$ & $-\frac{n-2}{n-3}$  & $0$ & $1< n\leq 2.5$&{\text{saddle}} \\ \hline
  $P_3^+$ & $\frac{1}{2-n}$ & $0$ & $-\frac{n-1}{2-n}$ & $\frac{n-1}{n(n-2)^2}$
  & $1< n\leq n_+$& {\text{non-hyperbolic with $3D$ stable manifold\ \ \  }}\\ \hline
    $P_3^-$ & $-\frac{1}{2-n}$ & $0$ & $\frac{n-1}{2-n}$ & $\frac{n-1}{n(n-2)^2}$
  & $1< n\leq n_+$& {\text{non-hyperbolic with $3D$ unstable manifold}} \\ \hline
  $P_4^+$ & $\frac{2n^2-5n+5}{7n^2-16n+10}$ & $-\frac{2n^2-2n-1}{7n^2-16n+10}$
  & $\frac{3(n-1)(n-2)}{7n^2-16n+10}$  & $\frac{9(4n^2-10n+7)(n-1)^2}{(7n^2-16n+10)^2}$ & $n_+\leq n\leq 3$
  & {\text{saddle with $3D$ stable manifold\ \ \  }}\\ \hline
  $P_4^-$ & $-\frac{2n^2-5n+5}{7n^2-16n+10}$ & $\frac{2n^2-2n-1}{7n^2-16n+10}$
  & $-\frac{3(n-1)(n-2)}{7n^2-16n+10}$  & $\frac{9(4n^2-10n+7)(n-1)^2}{(7n^2-16n+10)^2}$ & $n_+\leq n\leq 3$
  &{\text{saddle with $3D$ unstable manifold}} \\ \hline
  $P_5^+$ & $\frac{1}{n-2}$
  & $\frac{\sqrt{2n-5}}{n-2}$ & $\frac{n-3}{n-2}$ & $0$ & $2.5\leq n\leq 3$ &\text{non-hyperbolic}, with a 2D center manifold \\ \hline
  $P_5^-$ & $-\frac{1}{n-2}$
  & $\frac{\sqrt{2n-5}}{n-2}$ & $-\frac{n-3}{n-2}$ & $0$ & $2.5\leq n\leq 3$  &\text{non-hyperbolic}, with a 2D center manifold \\ \hline
  \hline
\end{tabular}
\caption{\label{tab1}The isolated critical points of the
cosmological system.
  We use the notation  $n_+=\frac{1+\sqrt{3}}{2}\approx 1.37$.}
\end{table*}

The eigenvalues of the Jacobian matrix are presented in Appendix
\ref{Eigenvalues} and thus here we straightway provide each
point's type.

\begin{itemize}
\item The two critical points ${\cal A_\epsilon}$ exist for
$-\frac{1}{6}\leq w\leq 1.$ The critical point ${\cal A}_+$
(${\cal A}_-$) is a sink, that is a stable one (source, that is an
unstable) provided $w_+<w\leq 1$, where $w_+=\frac{1}{3}
\left(-2+\sqrt{3}\right)\approx -0.09$. Otherwise they are saddle
points. Equivalently, we can express the stability intervals in
terms of $n$, using the relation (\ref{wn}). Thus, the
aforementioned interval becomes $n_+<n\leq 3$, where
$n_+=3(w_++1)/2\approx 1.37$.
 \item The critical points
${\cal B}_\epsilon$ exist for $-\frac{1}{3}\leq w\leq
\frac{2}{3}.$   Thus, they are always saddle points having a
2-dimensional stable manifold and a 2-dimensional unstable
manifold if $-\frac{1}{3}< w< \frac{2}{3}.$
 \item The critical points
$P_3^\epsilon$ exist for $-\frac{1}{3}\leq w\leq w_+.$ The are
non-hyperbolic, and due to the 1-dimensional center manifold
presented in the Appendix \ref{Eigenvalues}, the stable (unstable)
manifold of $P_3^+$ ($P_3^-$) is 3D
   for $-\frac{1}{3}<w<w_+.$
\item
The critical points $P_4^\epsilon$ exist for $w_+\leq w\leq
1.$
   $P_4^+$ ($P_4^-$) has a 3D  stable (unstable)
   manifold and a 1D  unstable (stable) manifold provided $w_+<w\leq
   1.$ Thus, they are always saddle points.
\item The critical points $P_5^\epsilon$ exist for
$\frac{2}{3}\leq w\leq 1.$  They are non-hyperbolic, there exists
a 2D center manifold and $P_5^+$ ($P_5^-$) has a 2D unstable
(stable) manifold.
\end{itemize}

Lastly, as we have mentioned, in this subsection we have
investigated the system \eqref{eqQ}-\eqref{eqy}, leaving outside
the decoupled equation \eqref{Aux}. However, for completeness, we
study equation \eqref{Aux}, and especially the stability across
the direction $\Ex$, in Appendix \ref{subsE1}.

\subsection{Physical description of the solutions and connection with observables}
\label{conobserv}

Let us now present the formalism of obtaining the physical
description of a critical point, and also connecting with the
basic observables relevant for a physical discussion. These will
allow us to describe the cosmological behavior of each critical
point, in the next section.

Firstly, around a critical point we obtain first-order expansions
for $\ex,{{}^2\!K}$, $\rho_m$ and $R$ in terms of $\tau$,
considering the versions of equations \eqref{aux}, \eqref{propK}
and \eqref{cosmeq5}, respectively given by
\begin{eqnarray}
\left(\ex\right)'&=&-(n-1)\left[Q^*-2\Sigma^*\right] \ex,\nonumber\\
\left({{}^2\!K}\right)'&=&-2(n-1)\left[Q^*+\Sigma^*\right]{{}^2\!K},\nonumber\\
\rho_m'&=&-2 n(n-1)Q^*\rho_m,\nonumber\\
R'&=&2 x^* R,\label{APPROX}
\end{eqnarray}
where the star-upperscript denotes the evaluation at a specific
critical point, and the prime denotes derivative with respect to
$\tau$. The last equation follows from the definition of $x$ given
by \eqref{auxiliary}. In general we can consider the case $x^*\neq
0$ and $y^\star\neq 0$, since in the simple case of $x^*=0$ and
$y^\star\neq 0$ we obtain $R=R_0$, with $R_0$ a constant, while in
the case
  $y^\star=0$ we acquire $R=0$.

In order to express the above determined functions of $\tau$ in
terms of the comoving time variable $t$, we invert the solution of
\begin{equation}
\frac{ {d }t}{ {d}\tau}=\frac{n-1}{D^*} \label{taueq}
\end{equation}
with $D^*$ being the first-order solution of
\begin{equation}
 D'=D (n-1) \Upsilon^\star  \label{evolD},
   \end{equation}
   where
\begin{eqnarray}
\Upsilon^\star= x^*-\Sigma^* +(Q^*+x^*) \left[-3 {x^*}^2+\Sigma^*
x^*\ \ \ \ \ \ \ \ \ \right.\nonumber\\
\left.\ \ \ \ \ \ \ \ \ +(Q^*-3 \Sigma^* ) \Sigma^* +n
\left({x^*}^2+{\Sigma^*}
   ^2+y^*-1\right)\right].
   \end{eqnarray}
Solving equations \eqref{taueq}-\eqref{evolD} (with initial
conditions $D(0)=D_0$ and $t(0)=t_0$) we obtain
\begin{equation}
t(\tau )=\frac{1-e^{(1-n) \tau  \Upsilon^* }}{D_0 \Upsilon^*
}+t_0.
\end{equation}
Thus, inverting the last equation for $\tau$ and substituting in
the solution of \eqref{APPROX} with initial conditions
$\ex(0)=\ex_0, {{}^2\!K}(0)={{}^2\!K}_0, \rho_m(0)=\rho_{m0},
R(0)=R_0$, we acquire
\begin{eqnarray}
&&\ex(t)=\ex_0 (D_0 (t_0-t) \Upsilon^* +1)^{\frac{Q^*-2 \Sigma^*
}{\Upsilon^* }},\nonumber\\
&&{{}^2\!K}(t)={{}^2\!K}_0 (D_0 (t_0-t) \Upsilon^*
+1)^{\frac{2 (Q^*+\Sigma^* )}{\Upsilon^* }}\nonumber\\
&&\rho_m(t)=\rho_{m0} (D_0 (t_0-t) \Upsilon^* +1)^{\frac{2 n
Q^*}{\Upsilon^* }},
\nonumber\\
&&R(t)=R_0 (D_0 (t_0-t) \Upsilon^* +1)^{\frac{2
x^*}{(1-n)\Upsilon^*
   }}.
   \label{physsolutions}
\end{eqnarray}
Finally, it can be shown that the length scale $\ell$
 along the flow lines, defined in (\ref{thetaelldef}), can be expressed as
\cite{EU}
\begin{equation}
\label{lengthscale}
\ell=\ell_0 (\ell_1 -t\Upsilon^*)^{-\frac{Q^*}{\Upsilon^* }},
\end{equation}
  where $\ell_0=\left[\left(\ex_0\right)\left({{}^2\!K}_0\right)\right]^{-\frac{1}{3}}$ and
  $ \ell_1=D_0 t_0 \Upsilon^*
  +1$.
In summary, expressions (\ref{physsolutions}) and
(\ref{lengthscale}) determine the solution, that is the evolution
of various quantities, at a critical point.

Let us now come to the observables. Using the above expressions,
we can calculate the deceleration parameter $q$ defined as usual
as \cite{EU}
\begin{equation}
\label{qq}q=-\frac{\ell \ddot\ell}{(\dot\ell)^2}.
\end{equation}
Additionally, we can calculate the effective (total)
equation-of-state parameter of the universe $w_{\text{eff}}$,
defined conventionally as
\begin{equation}
\label{weff}
w_{\text{eff}}\equiv\frac{P_{\text{tot}}}{\rho_{\text{tot}}}=\frac{\frac{p_m}{f'(R)}+P}{\frac{\rho_m}{f'(R)}+\mu},
\end{equation}
where $P_{\text{tot}}$ and $\mu_{\text{tot}}$ are respectively the
total isotropic pressure and the total energy density as they can
be read from equation \eqref{varyingaction}, where $P$ and $\mu$
are given by \eqref{matterfields1}. Therefore, in terms of the
auxiliary variables we have
\begin{eqnarray}
\label{qq2}
q&=& \frac{\Sigma ^2-x (2 Q+x)+1}{Q^2}-\frac{n y}{(n-1) Q^2}\\
w_{\text{eff}}&=&\frac{2 n y}{3 (n-1) \left(x^2+2 Q x+\Sigma
^2-1\right)}+\frac{1}{3}. \label{weff2}
\end{eqnarray}
Finally, the various density parameters defined in
(\ref{Omegas1})-(\ref{Omegas4}), in terms of the auxiliary
variables straightforwardly read
\begin{align}
&\Omega_k=\frac{\left(Q+x\right)^2-1}{Q^2}\nonumber\\
& \Omega_{m}=\frac{1-x^2-y-\Sigma^2}{Q^2}\nonumber\\
&\Omega_{\text{curv.fl}}=\frac{y-2 x Q}{Q^2}\nonumber\\
&\Omega_\sigma=\left(\frac{\Sigma}{Q}\right)^2\label{Omegas}.
\end{align}

In conclusion, at each critical point we can calculate the values
of the basic observables $q$, $w_{\text{eff}}$ and the various
density parameters, and also calculate the specific physical
solution, that is obtain the behavior of $\ell(t)$, $\rho_m(t)$
and $R(t)$.

Finally, it is interesting to notice that in Kantowski-Sachs
geometry one can easily handle isotropization. In particular, the
geometry becomes isotropic if $\sigma_+$ becomes zero, as can be
seen from (\ref{metric}) and (\ref{sigmaplus}). Thus, critical
points with $\Sigma=0$ (or more physically $\Omega_\sigma=0$)
correspond to Friedmann points, that is to isotropic universes,
and when such an isotropic point is an attractor then we obtain
asymptotic isotropisation in the future \cite{Goheer:2007wu}.

\section{Cosmological Implications}\label{sectionIV}

In the previous sections we formulated $f(R)$-gravity in the case
of the homogeneous and anisotropic Kantowski-Sachs geometry,
focusing on the $f(R)=R^n$ ansatz, and we performed a detailed
phase-space analysis. Thus, in the present section we discuss the
physical implications of the mathematical results, focusing on the
physical behavior and on observable quantities.

\begin{table*}
    \centering
\begin{tabular}{|c|c|c|c|c|c|c|}
  \hline   \hline
  Cr.P.  & $w_{\text{eff}}$ & $ q$  & $\Omega_k$ & $\Omega_m$ & $\Omega_{\text{curv.fl}}$ & $\Omega_\sigma$\\
  \hline \hline
  $\mathcal{A}_{\pm}$  & $-\frac{6 n^2-7 n-1}{3 (n-1) (2 n-1)}$ & $-\frac{2 n^2-2 n-1}{(n-1) (2
  n-1)}$ & 0 & 0 & 1 & 0

 \\ \hline
  ${\cal B}_\pm$ & $\frac{1}{3}$ & $1$
  & $0$ & $5-2n$& $2(n-2)$ & 0
   \\ \hline
  $P_1^\pm, P_2^\pm$  & $\frac{1}{3}$ & $ 2$  & 0 & 0 & 0 & 1
 \\ \hline
 $P_3^\pm$  & $-\frac{1}{3}$ & 0 & 0  & $-2 n+2+\frac{1}{n}$ & $2
 n-1-\frac{1}{n}$ & 0
 \\ \hline
  $P_4^\pm$  & $\frac{n (4 (9-4 n) n-21)-1}{3 n (8 (n-3) n+21)-3}$ &
$\frac{6-3 n}{n (2 n-5)+5}-1$ &  $-\frac{3 \left(2 n^2-2
n-1\right) \left(4 n^2-10 n+7\right)}{\left(2 n^2-5 n+5\right)^2}$
& 0 &$\frac{3 (n-1) \left(8 n^3-24
   n^2+21 n-1\right)}{\left(2 n^2-5 n+5\right)^2}$ & $\frac{\left(2 n^2-2 n-1\right)^2}{\left(2 n^2-5
   n+5\right)^2}$
 \\ \hline
  $P_5^\pm$ & $\frac{1}{3}$ & $2 (n-2)$  & 0 & 0 & $2(3-n)$ & $2n-5$
 \\ \hline
$C_\pm$ & $\frac{1}{3}$ & $\frac{2  }{1 \pm\sin (u)}$ & 0 & 0
&$\frac{2 \sin (u)}{\sin (u)\pm 1}$ & $\frac{\cos ^2(u)}{(\sin
(u)\pm 1)^2}$
\\   \hline
\end{tabular}
\caption{  \label{tab2a} The values of the basic observables,
namely the effective (total) equation-of-state parameter
$w_{\text{eff}}$, the deceleration parameter $q$, the curvature
density parameter $\Omega_k$, the matter density parameter
$\Omega_m$, the ``curvature-fluid'' density parameter
$\Omega_{\text{curv.fl}}$ and the shear density parameter
$\Omega_\sigma$, defined in (\ref{qq2})-(\ref{Omegas}), at the
isolated critical points and curves of critical points of the cosmological
system. We display the information about the non-isolated critical points $P_1^\pm, P_2^\pm$ to emphasize they are solutions dominated by shear.}
\end{table*}

\begin{table*}
    \centering
\begin{tabular}{ccc}
  \hline   \hline
  Cr.P.    & $\Upsilon^*$ & Solution/description\\
  \hline \hline
  $\mathcal{A}_+$    & $\frac{(n-2)  }{3
(n-1)^2}$ & $\ell(t)=\left\{\begin{array}{cc}\ell_0(\ell_1-t
\Upsilon^*)^{s_1} & n\neq 2\\
                                       \ell_0 e^{H_1 (t-t_0)} &
n=2\end{array}\right.,$ $\rho_m(t)=\rho_{m0}
\left[\frac{\ell}{\ell_0}\right]^{-2 n},$  $R(t)=\left\{\begin{array}{cc}\frac{R_0}{(\ell_1-t \Upsilon^*)^2}& n\neq \{1.25,2\}\\
        R_0& n=2\\
        0 & n=1.25\end{array}\right.$
\\\vspace{0.2cm}
           & & Isotropic. Expanding.  Accelerating for $n_+<n<3$. Phantom for $2<n<3.$
\\\vspace{0.2cm}
            & & For $n=2$ exhibits dS behavior
 \\ \hline
  $\mathcal{A}_-$   & $-\frac{(n-2)  }{3
(n-1)^2}$ & $\ell(t)=\left\{\begin{array}{cc}\ell_0(\ell_1-t
\Upsilon^*)^{s_1} & n\neq 2\\
                                       \ell_0 e^{H_1 (t-t_0)} &
n=2\end{array}\right.,$ $\rho_m(t)=\rho_{m0}
\left[\frac{\ell}{\ell_0}\right]^{-2 n},$  $R(t)=\left\{\begin{array}{cc}\frac{R_0}{(\ell_1-t \Upsilon^*)^2}& n\neq \{1.25,2\}\\
        R_0& n=2\\
        0 & n=1.25\end{array}\right.$
\\\vspace{0.2cm}
           & & Isotropic. Contracting. Decelerating for $n_+<n<3$. Phantom for $2<n<3.$
\\\vspace{0.2cm}
            & & For $n=2$ exhibits collapsing AdS behavior
 \\ \hline
  ${\cal B}_+$
  & $\frac{2  }{n-3}$ & $\ell(t)=\ell_0\sqrt{(\ell_1-t \Upsilon^*)},$
$\rho_m(t)=\rho_{m0} (\ell_1-t \Upsilon^*)^{-n},$ $R(t)=0$
\\\vspace{0.2cm}
       & & Isotropic. Expanding. Decelerating. Total matter/energy mimics
      radiation.
   \\ \hline
    ${\cal B}_-$
  & $-\frac{2  }{n-3}$ & $\ell(t)=\ell_0\sqrt{(\ell_1-t \Upsilon^*)},$
$\rho_m(t)=\rho_{m0} (\ell_1-t \Upsilon^*)^{-n},$ $R(t)=0$
\\\vspace{0.2cm}
      & &  Isotropic. Contracting. Accelerating. Total matter/energy mimics radiation.
 \\ \hline
  $P_3^+$   & $-\frac{1 }{n-2}$ &
$\ell(t)=\ell_0{(\ell_1-t \Upsilon^*)}=a_1+a_2 t,$
$\rho_m(t)=\rho_{m0} (\ell_1-t \Upsilon^*)^{-2n}$,
$R(t)=\frac{R_0}{(\ell_1-t \Upsilon^*)^2}$ \\\vspace{0.2cm}
     & & Isotropic. Expanding. Zero acceleration.
 \\ \hline
  $P_3^-$   & $\frac{1 }{n-2}$ &
$\ell(t)=\ell_0{(\ell_1-t \Upsilon^*)}=a_1+a_2 t,$
$\rho_m(t)=\rho_{m0} (\ell_1-t
\Upsilon^*)^{-2n}$,$R(t)=\frac{R_0}{(\ell_1-t \Upsilon^*)^2}$
\\\vspace{0.2cm}
    & & Isotropic. Contracting. Zero acceleration.
 \\ \hline
  $P_4^+$  & $\frac{3 (n-2)  }{7 n^2-16 n+10}$
& $\ell(t)=\left\{\begin{array}{cc}\ell_0(\ell_1-t
\Upsilon^*)^{s_2} & n\neq 2\\
                                       \ell_0 e^{H_2 (t-t_0)} &
n=2\end{array}\right.,$ $\rho_m(t)=\rho_{m0}
\left[\frac{\ell}{\ell_0}\right]^{-2 n},$
$R(t)=\left\{\begin{array}{cc}\frac{R_0}{(\ell_1-t
\Upsilon^*)^2}& n\neq 2\\
R_0, & n=2\end{array}\right.$ \\\vspace{0.2cm}
                                    & &Expanding. Accelerating for
$n_+<n<3$. Phantom for $2<n<M_+$.
\\\vspace{0.2cm}
            & & For $n=2$ exhibits dS behavior
 \\ \hline
 $P_4^-$  & \ $-\frac{3 (n-2)  }{7 n^2-16
n+10}$ & $\ell(t)=\left\{\begin{array}{cc}\ell_0(\ell_1-t
\Upsilon^*)^{s_2} & n\neq 2\\
                                       \ell_0 e^{H_2 (t-t_0)} &
n=2\end{array}\right.,$ $\rho_m(t)=\rho_{m0}
\left[\frac{\ell}{\ell_0}\right]^{-2 n},$
$R(t)=\left\{\begin{array}{cc}\frac{R_0}{(\ell_1-t
\Upsilon^*)^2}& n\neq 2\\
R_0, & n=2\end{array}\right.$ \\\vspace{0.2cm}
                                      & &Contracting. Decelerating for
$n_+<n<3$. Phantom for  $2<n<M_+$.
\\\vspace{0.2cm}
          & & For $n=2$ exhibits collapsing AdS behavior
 \\ \hline
  $P_5^+$   & $\frac{(3-2 n)
}{n-2}$ &  $\ell(t)=\ell_0(\ell_1-t \Upsilon^*)^{s_3},$
$\rho_m(t)=\rho_{m0} \left[\frac{\ell}{\ell_0}\right]^{-2 n}$,
$R(t)=0$
   \\\vspace{0.2cm}
                  & & Expanding.
                  Decelerating. Total matter/energy mimics
      radiation.
 \\ \hline
  $P_5^-$   & $-\frac{(3-2 n)
}{n-2}$ &  $\ell(t)=\ell_0(\ell_1-t \Upsilon^*)^{s_3},$
$\rho_m(t)=\rho_{m0} \left[\frac{\ell}{\ell_0}\right]^{-2 n}$,
$R(t)=0$
 \\\vspace{0.2cm}
                  & & Contracting. Accelerating. Total matter/energy mimics
      radiation.\\
  \hline
\end{tabular}
\caption{  \label{tab2b} The behavior of  $\ell(t)$ (length scale
along the flow lines), of $\rho_m(t)$ (matter energy density) and
of $R(t)$ (Ricci scalar) at the critical points of the
cosmological system. We use the notations $s_1=-\frac{(n-1) (2
n-1)}{n-2},$ $s_2=-\frac{2 n^2-5 n+5}{3 (n-2)},$ $s_3=\frac{1}{2
n-3},$ $p_\epsilon(u)=\frac{\epsilon +\sin (u)}{3 \epsilon +\sin
(u)},$ $n_+=\frac{1+\sqrt{3}}{2}\approx 1.37$ and $M_+=\frac{1}{4}
\left(5+\sqrt{21}\right)\approx 2.40.$ }
\end{table*}

\begin{table*}
    \centering
\begin{tabular}{ccc}
  \hline   \hline
  \ \ Cr.P.  \ \    & $\Upsilon^*$ & Solution/description\\
  \hline \hline
  $C_+$ &
 $-\left[3  +\sin (u)\right]$ & $\ell(t)=\ell_0(\ell_1-t
\Upsilon^*)^{p_+(u)},$ $\rho_m(t)=\rho_{m0}
\left[\frac{\ell}{\ell_0}\right]^{-2 n}$, $R(t)=0$\\\vspace{0.2cm}
   & & Expanding. Decelerating. Total matter/energy mimics
      radiation.\\
  \hline
  $C_-$ &  \ \ \ \ \ $-\left[-3  +\sin (u)\right]$ \ \ \ \ \ &
$\ell(t)=\ell_0(\ell_1-t \Upsilon^*)^{p_-(u)},$
$\rho_m(t)=\rho_{m0} \left[\frac{\ell}{\ell_0}\right]^{-2 n}$,
$R(t)=0$\\\vspace{0.2cm}
   & &  Contracting. Accelerating. Total matter/energy mimics
      radiation.\\
  \hline
\end{tabular}
\caption{ \label{tab2c} Physical behavior of the solutions at the
curves of critical points of the cosmological system. We use the
notation $p_\epsilon(u)=\frac{\epsilon +\sin (u)}{3 \epsilon +\sin
(u)}$.
 }
\end{table*}

First of all, for each critical point of Tables \ref{tab1b} and
\ref{tab1} we calculate the effective (total) equation-of-state
parameter of the universe $w_{\text{eff}}$ using (\ref{weff2}),
and the deceleration parameter $q$ using (\ref{qq2}), and the
values of the density parameters $\Omega_k,$  $\Omega_m,$
$\Omega_{\text{curv.fl}},$ and $\Omega_\sigma$ using
\eqref{Omegas}. These results are presented in the Table
\ref{tab2a}.

Furthermore, for each critical point we use  (\ref{physsolutions})
and (\ref{lengthscale}), in order to extract the behavior of the
physically important quantities  $\ell(t)$, $\rho_m(t)$ and $R(t)$
at this critical point. As we have mentioned
 $\ell(t)$ is the length scale along the flow lines and in the case of zero
anisotropy (for instance in FRW cosmology) it is just the usual
scale factor. Additionally, $\rho_m(t)$ is the matter energy
density and $R(t)$ is the Ricci scalar. These solutions are
presented in the last column of Tables \ref{tab2b} and
\ref{tab2c}.

Finally, in the last column of Tables \ref{tab2b} and \ref{tab2c}
we also present the physical description of the corresponding
solution, taking into account all the above information. In
particular, since the auxiliary variable $Q$ defined in
(\ref{auxiliary}) is the Hubble scalar divided by a positive
constant, $Q>0$ corresponds to an expanding universe, while $Q<0$
to a contracting one. Furthermore, as usual, for an expanding
universe $q<0$ corresponds to accelerating expansion and $q>0$ to
decelerating expansion, while for a contracting universe $q<0$
corresponds to decelerating contraction and $q>0$ to accelerating
contraction. Additionally, if $w_{\text{eff}}<-1$ then the total
equation-of-state parameter of the universe exhibits phantom
behavior. Lastly, critical points with $\Sigma=0$ correspond to
isotropic universes.

Let us analyze the physical behavior in more details. The critical
point  $\mathcal{A}_+$ is stable for $1.37\lesssim n<3$, and thus
at late times the universe can result in it. It corresponds to an
isotropic expanding universe, in which the expansion is
accelerating. Additionally, since the curvature energy density is
zero, this point corresponds to an asymptotically flat universe
(this is equivalent to close or open FRW universes which may
possess zero curvature solutions as asymptotic states).
Furthermore, in the case $2<n<3$ the total equation-of-state
parameter of the universe exhibits phantom behavior. Thus, the
anisotropic Kantowsky-Sachs $R^n$-gravity can lead the universe
not only to be accelerating, but also to be in the phantom
``phase'', a result of great cosmological interest. We have to
mention here that since point $\mathcal{A}_+$ is asymptotically an
FRW one, one expects to obtain phantom behavior in isotropic (FRW)
$R^n$-gravity too. Although in the phase-space analysis of FRW
$R^n$-gravity \cite{goheer}, the authors have focused only on
acceleration, without examining the total equation-of-state
parameter of the universe, it is easy to see that if ones defines
it and examines its features he will find phantom behavior in that
case too. This is also the case in Bianchi I and Bianchi III
$R^n$-gravity \cite{Goheer:2007wu,Shamir:2010ee}, where the
authors would have found the phantom behavior if they had
calculated the total equation-of-state parameter. Therefore, we
conclude that the phantom behavior is a result of the modification
of gravity, as it has been discussed in detail in the literature
\cite{Sotiriou:2008rp,DeFelice:2010aj,Nojiri:2003ft}. Finally, in
the case where the matter is dust ($w=0$, that is $n=3/2$),
$\rho_m(t)$ behaves like $\ell(t)^{-3}$ as expected, and this acts
as a self-consistency test for our analysis. Additionally, note
that in the special case $n=2$, that is when $w=1/3$, i.e when
radiation dominates the universe, the aforementioned stable
solution corresponds to a de-Sitter expansion (this is not a new
feature since de-Sitter solutions are known to exist in Bianchi I
and Bianchi III $R^n$-gravity \cite{Goheer:2007wu}). This is of
great significance since such a behavior can describe the
inflationary epoch of the universe.

In the above critical point the isotropization has been achieved.
Such late-time isotropic solutions, than can attract an initially
anisotropic universe, are of significant cosmological interest and
have been obtained and discussed in the literature
\cite{Leach:2006br,Goheer:2007wu}. The acquisition of such a
solution was one of the motives of the present work, as of many
works on anisotropic cosmologies, since, as we discussed in the
Introduction, it is the only robust approach in confronting
isotropy of standard cosmology. The fact that this solution is
accompanied by acceleration or phantom behavior, makes it a very
good candidate for the description of the observable universe.

The critical point  $\mathcal{A}_-$ which corresponds to an
isotropic contracting universe is not stable and thus it cannot be
the late-time state of the universe. Similarly, the points ${\cal
B}_+$ and   ${\cal B}_-$, which correspond to isotropic expanding
and contracting universes respectively, and in which the total
matter/energy mimics radiation, are saddle points and thus they
cannot be the late-time solution for the universe, too. The points $P_5^\epsilon$, which correspond to
decelerating expansion (for $\epsilon=+1$) and to accelerating
contraction (for $\epsilon=-1$), are non-hyperbolic with a
2D center manifold, and thus, generically,
the universe cannot be led to them.

The critical points $P_1^-$ and $P_2^-$ which correspond to
accelerating contraction, in which the total matter/energy behaves
like radiation, are non-hyperbolic critical points, but they do
possess a $3D$ stable manifold. By an explicit and straightforward
computation of their center manifolds \cite{wiggins} we deduce
that for $2<n<3$ each center manifold is locally asymptotically
stable (the equation governing the dynamics on the center manifold
is a gradient-type differential equation with potential function
having a degenerate local minimum at the origin), and thus $P_1^-$
and $P_2^-$ are locally asymptotically stable \cite{wiggins}, and
can attract the universe at late times. On the contrary, for
$1<n<2$, $P_1^-$ and $P_2^-$ are locally asymptotically unstable
(of saddle type). In summary, although
 $P_1^-$ and $P_2^-$ are not stable, they do have a
significant probability to be a late-time state for the universe
(this is realized for initial conditions on its stable manifold),
or at least the universe can stay near these solutions for a long
time before the dynamics remove it from them. On the other hand,
the points $P_1^+$ and $P_2^+$ which correspond to decelerating
expansion, possess a $3D$ unstable manifold, and thus they cannot
be a late-time solution of the universe.

Non-hyperbolic critical point with a $3D$ stable manifold is also the critical
point $P_3^+$, which corresponds to an asymptotically flat
isotropic expansion with zero acceleration, and thus it has also a
large probability to be a late-time state of the universe.
However, note that this solution corresponds to zero acceleration,
and thus $\ell(t)$ is a linear function of $t$. On the other hand,
the critical point $P_3^-$ (isotropic, contracting with zero
acceleration) possesses a $3D$ unstable manifold and thus it
cannot attract the universe at late times.


The critical point $P_4^+$ is saddle with a $3D$ stable manifold,
and thus it has a large probability to be a late-time state of the
universe. It corresponds to a non-flat, accelerating expansion,
for $1.37\lesssim n<3$, and furthermore in the case $2<n<3$ it
exhibits phantom behavior. Finally, for $n=2$, that is for
$w=1/3$, it corresponds to a de-Sitter expansion. On the other
hand $P_4^-$ (decelerating contraction) is highly unstable and
therefore it cannot attract the universe at late times.

We mention here that the critical points
$P_1^\epsilon$,$P_2^\epsilon$,$P_3^\epsilon$,$P_4^\epsilon$ and
$P_5^\epsilon$ are not present in isotropic (FRW) $R^n$-gravity,
as  compared with \cite{goheer}. They arise as a pure result of
the anisotropy, and this shows that the much more complicated
structure of anisotropic geometries leads to radically different
cosmological behaviors comparing to the simple isotropic
scenarios.

Additionally, we have to analyze the behavior of the curves of
critical points $C_\epsilon$. The points $C_-$, which correspond
to accelerating contraction, are stable if $1.25\leq n\leq2.5$,
and thus they can be late-time solutions of the universe. On the
other hand, $C_+$, which correspond to decelerating expansion, are
unstable and thus they cannot attract the universe at late times.

Let us finish the physical discussion by referring to static
solutions, in order to compare to the FRW case of \cite{goheer}.
From the cosmological point of view, static solutions possess
$\ell(t)=const.$, that is $\dot{\ell}(t)=0$ and
$\ddot{\ell}(t)$=0, in order to obtain $H(t)=0$ and $\dot{H}(t)=0$
(note that one needs both conditions, since in a cosmological
bounce or turnaround, that is when a universe changes from
contracting to expanding or vice versa, $\dot{\ell}(t)$ is zero
instantly, but then it becomes positive or negative again). Thus,
using our auxiliary variable $Q$ defined in (\ref{auxiliary})
static solutions should have $Q=0$ and $\dot{Q}=0$. In conclusion,
since in all the aforementioned critical points $\dot{Q}=0$ by
definition, static solutions are just those with $Q=0$. At this
point there is another important difference comparing to the
isotropic case of \cite{goheer}. In particular, in FRW geometry,
the Ricci scalar is $R=6(\dot{H}+2H^2)$, and therefore static
solutions have $R=0$ (however the inverse is not true, that is not
all solutions with $R=0$ correspond to static ones, since one can
have $R=0$ but with $\dot{H}$ and $H$ non-zero). On the other
hand, in the anisotropic case $R$ is given by (\ref{Riccidef}),
i.e $  R= 12H ^{2}+6 \sp^{2}+6\dot H +2{}^2\!K$, that is we have
also the presence of additional terms, and therefore  static
solutions do not correspond to $R=0$ unless $\sigma_+$ and $^2K$
are also zero, which is not fulfilled in general. Thus, in the
anisotropic case one cannot use $R$, or equivalently the auxiliary
variable $y$ defined in (\ref{auxiliary}), in order to straightway
determine the static solutions, in contrast to the isotropic case
\cite{goheer} where the authors use $y=0$ for such a
determination.

\begin{figure}[ht]
\begin{center}
\includegraphics[width=9cm, height=8cm,angle=0]{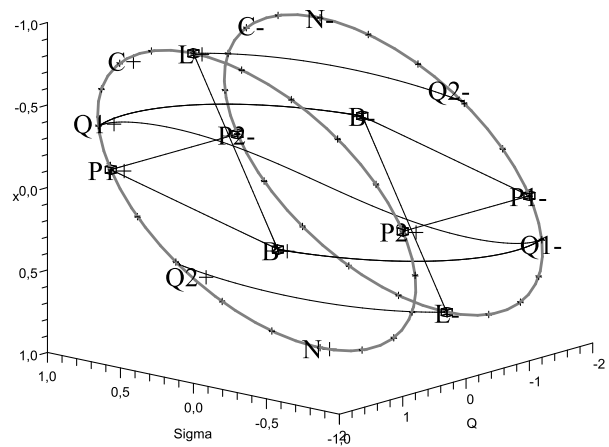}
\caption{{\label{fig1}} Projection of the phase space  on the
invariant set $y=0$,  in the case of dust matter ($w=0,n=3/2$).
 The critical points $Q_{1,2}^\epsilon$ have
coordinates $Q_1^\epsilon:=(Q=4\epsilon/3,
\Sigma=2\sqrt{2}\epsilon/3,-\epsilon/3)$ and
$Q_2^\epsilon:=(Q=\epsilon/3,
\Sigma=\sqrt{5}\epsilon/3,2\epsilon/3).$ All the points in the
circle $C_+$ are unstable whereas all the points in the circle
$C_-$ are stable. The heteroclinic sequences reveal the possible
transition from expansion to contraction and vice versa (see
text).}
\end{center}
\end{figure}

Now, in order to present the aforementioned results in a more
transparent way, we perform a numerical elaboration of our
cosmological system, using a seventh-eighth order continuous
Runge-Kutta method with absolute error $10^{-8}$, and relative
error $10^{-8}$ \cite{maple}. In Fig. \ref{fig1} we depict some
orbits in the invariant set $y=0$, in the case of dust matter
($w=0,n=3/2$).
  As we observe, we have the appearance of four
critical points with coordinates $Q_1^\epsilon:=(Q=4\epsilon/3,
\Sigma=2\sqrt{2}\epsilon/3,-\epsilon/3)$ and
$Q_2^\epsilon:=(Q=\epsilon/3,
\Sigma=\sqrt{5}\epsilon/3,2\epsilon/3)$, located in the invariant
curves $C^\epsilon$. All the points in the circle $C_+$ are
unstable whereas all the points in the circle $C_-$ are stable.
Note that the critical points $P_3^\epsilon$ coincide with ${\cal
N}_\epsilon$. In the figure we also  display heteroclinic
sequences \cite{REZA} of types
\begin{eqnarray}
&& Q_1^+ \longrightarrow \left\{\begin{array}{c}
  {\cal B}^-\longrightarrow {\cal L}_- \\
  {\cal B}^-\longrightarrow P_1^-  \\
  Q_1^- \ \ \ \ \ \ \ \ \ \ \ \ \\
\end{array}\right.\nonumber\\
&&{\cal L}_+ \longrightarrow \left\{\begin{array}{c}
    {\cal B}^+\longrightarrow Q_1^-  \\
  Q_2^-\ \ \ \ \ \ \ \ \ \ \ \  \\
\end{array}\right.\nonumber\\
&&P_1^+ \longrightarrow \left\{\begin{array}{c}
    {\cal B}^+\longrightarrow Q_1^-  \\
  P_2^- \ \ \ \ \ \ \ \ \ \ \ \ \\
\end{array}\right.\nonumber\\
&&P_2^+ \longrightarrow P_1^-\nonumber\\
&&Q_2^+ \longrightarrow {\cal L}_-.
\end{eqnarray}
Thus, we can have evolutions in which the $+$ branch and $-$
branch are connected, that is we can have the transition from
expansion to contraction and vice versa. This is just a
cosmological turnaround and a cosmological bounce, and their
realization in the present scenario reveals the capabilities of
the model. It is interesting to note that such behaviors can be
realized in FRW $R^n$-gravity \cite{Carloni05,Nojiri06}, however
they are impossible in General Relativity Kantowsky-Sachs
cosmology \cite{Solomons:2001ef}. Therefore, we conclude that they
are a result of the  $R^n$ gravitational sector and not of the
anisotropy.

\begin{figure}[!]
\begin{center}
\includegraphics[width=9cm, height=9cm,angle=0]{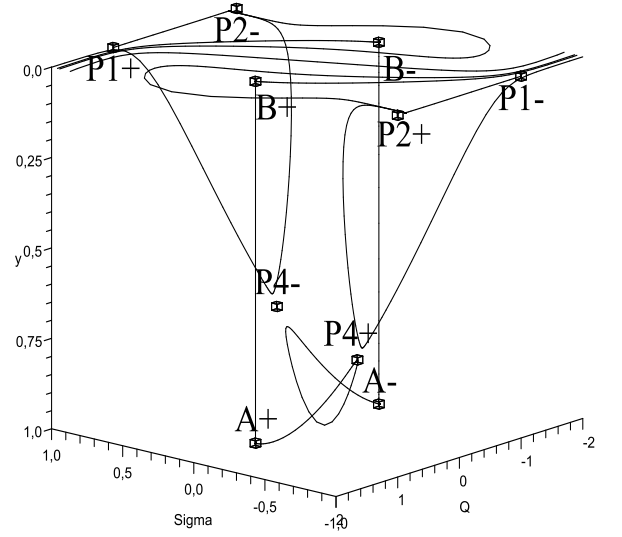}
\caption{{\label{fig2a}} Projection of the phase space on the
invariant set $x=0$, in the case of radiation ($w=1/3, n=2$).
There is one orbit of type ${\cal B}_+\longrightarrow {\cal A}_+$
and one of type ${\cal A}_-\longrightarrow {\cal B}_-$. The
existence of an heteroclinic sequence of type ${\cal
A}_-\longrightarrow P_4^-\longrightarrow P_4^+\longrightarrow
{\cal A}_+$ corresponds to a cosmological bounce.}
\end{center}
\end{figure}

In Fig. \ref{fig2a} we display some orbits in the case of
radiation ($w=1/3, n=2$), where as we mentioned in subsection
\ref{sectionIII.2} the invariant set $x=0$ appears. Particularly,
we observe the existence of heteroclinic sequences of types
\begin{eqnarray}
&&{\cal A}_-\longrightarrow \left\{\begin{array}{c}
  P_4^-\longrightarrow P_4^+\longrightarrow {\cal A}_+ \\
  {\cal B}^- \ \ \ \ \ \ \ \ \ \ \ \ \ \ \ \ \ \ \ \ \ \ \ \\
\end{array}\right.\nonumber\\
&&{\cal B}_+\longrightarrow {\cal A}_+\nonumber\\
 &&P_1^-\longrightarrow
P_4^+\longrightarrow P_2^+\nonumber\\
 &&P_2^-\longrightarrow
P_4^-\longrightarrow P_1^+\nonumber\\
 &&P_1^+\longrightarrow P_2^-\nonumber\\
&&P_2^+\longrightarrow P_1^-,\label{heteroclinic}
\end{eqnarray}
revealing the realization of a cosmological bounce or a
cosmological turnaround. Similarly to the isotropic case (see
figure 5 of \cite{goheer}) there is one orbit of type ${\cal
B}_+\longrightarrow {\cal A}_+$ and one of type ${\cal
A}_-\longrightarrow {\cal B}_-$. However, in the present case we
have the additional existence of an heteroclinic sequence of type
${\cal A}_-\longrightarrow P_4^-\longrightarrow
P_4^+\longrightarrow {\cal A}_+$, corresponding to the transition
from collapsing AdS to expanding dS phase, that is we obtain a
cosmological bounce followed by a de Sitter expansion, which could
describe the inflationary stage. This significant behavior is a
pure result of the anisotropy and reveals the capabilities of the
scenario.  Lastly, note the very interesting possibility, of the
eternal transition $P_1^-\longrightarrow P_2^+\longrightarrow
P_1^-\longrightarrow P_2^+\cdots$ which is just the realization of
cyclic cosmology \cite{cyclic}. Bouncing solutions are found to
exist both in FRW $R^n$-gravity \cite{Carloni:2005ii}, as well as
in the Bianchi I and Bianchi III $R^n$-gravity
\cite{Goheer:2007wu} (see also \cite{Barragan:2010qb}), and thus
they arise from the $R^n$ gravitational sector. However, in the
present Kantowsky-Sachs geometry cyclicity seems to be realized
relatively easily (however without accompanied by isotropization),
that is without fine-tuning the model-parameters, which is an
advantage of the scenario.

\begin{figure}[!]
\begin{center}
\includegraphics[width=8cm, height=8cm,angle=0]{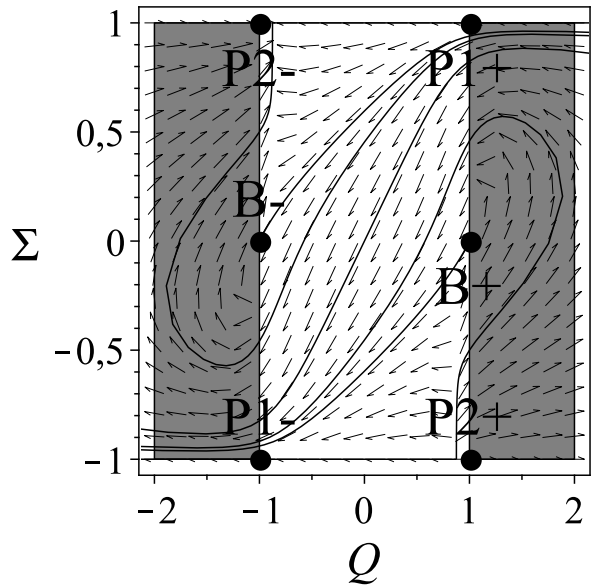}
\caption{{\label{fig2b}} Invariant set $x=0$, $y=0$ in the case of
radiation ($w=1/3, n=2$). The shaded region corresponds to the
unphysical portion of the phase plane.  $P_2^-$ ($P_2^+$) is the
local future (past) attractor, while $B_+, B_-$ and $P_1^+, P_1^-$
are saddle points.}
\end{center}
\end{figure}

Finally, in Fig. \ref{fig2b} we have extracted orbits located at
$x=y=0$, in the case of radiation.
The shaded region corresponds to the unphysical portion of the
phase plane (note however that this region is not invariant, since
an open set of orbits enter/abandon the unphysical boundary, and
thus such evolutions have to be excluded too). As we observe, in
this figure the last two heteroclinic sequences of
\eqref{heteroclinic} are displayed.

Let us make a comment here by referring to the cosmological epoch
sequence. As we mentioned in subsection \ref{sectionII.4}, in the
scenario at hand we can obtain the transition from the
matter-dominated era to the accelerated one. However, note that
the dynamical system analysis can only give analytical results
relating to the late-time states of the universe. The precise
evolution of the universe towards such late-time attractors
depends on the initial conditions, and it can only be investigated
through a detailed numerical elaboration similar to the partial
one that we performed in order to produce the aforementioned three
figures. Thus, by suitably determining the initial conditions, one
can obtain universes that start with a de-Sitter expansion, then
transit to the matter-dominated era, and finally falling into the
late time accelerating solution. The detailed examination of such
behaviors, along with the investigation of the basins of
attraction of the various evolutions in terms of the initial
conditions, and the estimation of the measures of the
corresponding sub-spaces of the phase space, lie outside of the
present work and are left for a future project.

We close this section by discussing the cosmological evolution in
the special case where $\Ex=0$ or $K=0$, since in this case
 the Kantowsky-Sachs metric
(in a comoving frame)
\begin{eqnarray}
 ds^2 = - \frac{(n-1)^2}{D^2}d\tau^2 + \frac{1}{ D^2\left(\Ex\right)^2}
 dr^2\ \ \ \ \ \ \ \ \ \, \ \ \ \ \nonumber\\
 \ \ \ \ \ \ + \frac{1}{3 D^2 K} [d\y^2 + \sin^2\varphi\,
   d\z^2],\label{dimensionlessmetric}
\end{eqnarray}
is singular. Although these points are unphysical their
neighbouring solutions could have physical meaning, which can be
extracted by obtaining first-order evolution rates valid in an
small neighborhood of the critical points of the system. In
particular, we first evaluate the perturbation-matrix ${\bf
{\Xi}}$ at the critical point of interest, we diagonalize it and
we obtain orders of magnitude for linear combinations of the
vector components $\left(\Ex,Q, \Sigma, x,y\right)^T$. Thus, the
desired expansion can be obtained by taking the inverse linear
transformation, and finally we preserve only the leading order
terms as $\tau\rightarrow -\infty$ or as $\tau\rightarrow \infty$.
Although this procedure is straightforward in the case of dust
matter ($w=0$, $n=3/2$) and radiation ($w=1/3$, $n=2$) we do not
present it explicitly since we desire to remain in the general
case of $\Ex\neq0$ and $K\neq0$.

\section{Conclusions}\label{conclusions}

In this work we constructed general anisotropic cosmological
scenarios where the gravitational sector belongs to the extended
$f(R)$ type, and we focused on Kantowski-Sachs geometries in the
case of $R^n$-gravity. We performed a detailed phase-space
analysis, extracting the late-time solutions, and we connected the
mathematical results with physical behaviors and observables.

As we saw, the universe at late times can result to a state of
accelerating expansion, and additionally, for a particular
$n$-range ($2<n<3$) it exhibits phantom behavior. Additionally,
the universe has been isotropized, independently of the anisotropy
degree of the initial conditions, and it asymptotically becomes
flat. The fact that such features are in agreement with
observations \cite{obs,Komatsu:2010fb} is a significant advantage
of the model. Moreover, in the case of radiation ($n=2$, $w=1/3$)
the aforementioned stable solution corresponds to a de-Sitter
expansion, and it can also describe the inflationary epoch of the
universe.

Note that at first sight the above behavior could be ascribed to
the cosmic no-hair theorem \cite{Wald}, which states that a
solution of the cosmological equations, with a positive
cosmological constant and under the perfect-fluid assumption for
matter, converges to the de Sitter solution at late times.
However, we mention that such a theorem holds for matter-fluids
less stiff than radiation, but more importantly it has been
elaborated for General Relativity \cite{Kitada-Maeda}, without a
robust extension to higher order gravitational theories
\cite{Cotsakis:1993er}. In our work we extracted our results
without relying at all on the cosmic no-hair theorem, which is a
significant advantage of the analysis.

Apart from the above behavior, in the scenario at hand the
universe has a large probability to remain in a phase of
(isotropic or anisotropic) decelerating expansion for a long time,
before it will be attracted by the above global attractor at late
times, and this acts as an additional advantage of the model,
since it is in agreement with the observed cosmological behavior.
However, the precise duration of such a  transient phase, unlike
the attractor behavior, does depend on the initial conditions of
the universe, which have to be suitable determined in order to
lead to a deceleration-duration of the order of $10^9$-$10^{10}$
years, before the universe pass to the accelerating phase. Such an
analysis can only arise through an explicit numerical elaboration,
that is beyond the analytical, dynamical-system treatment which is
where the present work focuses.

The Kantowski-Sachs anisotropic $R^n$-gravity can also lead to
contracting solutions, either accelerating or decelerating, which
are not globally stable. Thus, the universe can remain near these
states for a long time, before the dynamics remove it towards the
above expanding, accelerating, late-time attractors. The duration
of these transient phases depends on the specific initial
conditions.

One of the most interesting behaviors is the possibility of the
realization of the transition between expanding and contracting
solutions during the evolution. That is, the scenario at hand can
exhibit the cosmological bounce or turnaround. Additionally, there
can also appear an eternal transition between expanding and
contracting phases, that is we can obtain cyclic cosmology. These
features can be of great significance for cosmology, since they
are desirable in order for a model to be free of past or future
singularities.

Before closing, let us make some comments concerning the use of
observational data in order to constrain the present scenario. In
particular, equation (\ref{gauss111}), along with the
$R^n$-ansatz, relation (\ref{wn}), and the definitions of the
density parameters (\ref{Omegas1})-(\ref{Omegas4}), can be
straightforwardly written in the form used for observational
fitting \cite{Dutta:2009jn}. Thus, one can use observational data
from Type Ia Supernovae (SNIa), Baryon Acoustic Oscillations
(BAO), and Cosmic Microwave Background (CMB), along with
requirements of Big Bang Nucleosynthesis (BBN), to constrain the
model parameter $n$. Additionally, one can constrain the initial
allowed anisotropy value $\sigma_+$ (or $\Omega_\sigma$), through
its present value $\Omega_{\sigma0}$ and the shear evolution
(\ref{shear}). Such a procedure is necessary for every
cosmological paradigm, and can significantly enlighten the
scenario at hand. However, since in the present work we focused on
the dynamical features, the detailed observational elaboration is
left for a separate project.

In summary, anisotropic $R^n$-gravity has a very rich cosmological
behavior, and a large variety of evolutions and late-time
solutions, compatible with observations. The much more complicated
structure of anisotropic geometries leads to radically different
implications comparing to the simple isotropic scenarios. These
features indicate that anisotropic universes governed by modified
gravity can be a candidate for the description of nature, and
deserve further investigation.

\begin{acknowledgments}
The authors would like to thank P.K.S. Dunsby, S. Dutta, A.
Guarnizo, G. Kofinas, G.J. Olmo and A.A. Sen, for valuable
discussions, and to two anonymous referees for useful comments and
suggestions. G. L wishes to thank the MES of Cuba for partial
financial support of this investigation. His research was also
supported by Programa Nacional de Ciencias B\'asicas (PNCB).
\end{acknowledgments}

\appendix

\section{Eigenvalues of the perturbation matrix $\bf{\Xi}$ for the
critical points} \label{Eigenvalues}

The system (\ref{eqQ})-(\ref{eqy}) admits ten isolated critical
points presented in Table \ref{tab1}. Here we provide the
eigenvalues of the perturbation matrix $\bf{\Xi}$ calculated at
each critical point. We also provide the eigenvalues of the
perturbation matrix at the special points $P_1^\epsilon$ and
$P_2^\epsilon$ located at the invariant circles $C_\epsilon.$

For the two critical points ${\cal A_\epsilon}$ the associated
eigenvalues read
\begin{eqnarray}
\left\{-\epsilon\frac{2 \left(9 w^2+12 w+1\right)}{3 (3 w+1)},
-\epsilon\frac{2 (3 w+2) (6 w+1)}{3 (3 w+1)} [\times
2],\right.\nonumber\\
\left.-\epsilon\frac{(w+1) \left(9 w^2+12 w+1\right)}{3
w+1}\right\},\nonumber
\end{eqnarray}
 with $[\times 2]$ denoting multiplicity 2.

For the critical points ${\cal B}_\epsilon$ the associated
eigenvalues are
\[\left\{-\epsilon\frac{2 (3 w+1)}{3 (w-1)},-\epsilon\frac{(3 w-2)
(3 w+1)}{3 (w-1)} [\times 2],-\epsilon\frac{2
(w+1)}{w-1}\right\}.\]

For the critical points $P_1^\epsilon$ the associated eigenvalues
 read
\[\left\{0,\epsilon(3 w+1),3 \epsilon(3 w+1),-\frac{3}{2}\epsilon
(w-1) (3 w+1)\right\},\] while for $P_2^\epsilon$ they write
\[\left\{0,3 \epsilon(3 w+1) [\times 2],-\frac{3}{2} (w-1) (3
w+1)\right\}.\]

For the critical points $P_3^\epsilon$ the associated eigenvalues
are
\begin{eqnarray}
&&\left\{0,-\epsilon(3 w+1),\right.\nonumber\\
&&\ \ -\epsilon\frac{9 w^2+\sqrt{3} \sqrt{3 w+1} \sqrt{9 w^3+21
w^2+31
w+3}-1}{2 (3 w-1)},\nonumber\\
&&\ \ \left. -\epsilon\frac{9 w^2-\sqrt{3} \sqrt{3 w+1}
   \sqrt{9 w^3+21 w^2+31 w+3}-1}{2 (3 w-1)}\right\},\nonumber
\end{eqnarray}
 and thus there exists a 1-dimensional center
manifold tangent to the line
 \begin{eqnarray}
 &&\left\{x= -\frac{1}{2} Q (3 w+1),y= \epsilon\frac{Q (w-1) (3 w+1)^2}{(w+1) (3 w-1)},\right.\nonumber\\
 &&\ \  \left. \Sigma =-\frac{1}{2} Q (3 w-1), \, Q\in[-2,2]\right\}.\nonumber
\end{eqnarray}

For the critical points $P_4^\epsilon$ the associated eigenvalues
are
\begin{widetext}
\begin{eqnarray}
&&\left\{-\epsilon\frac{6 (3 w+1) \left(9 w^2+3 w+1\right)}{63
w^2+30 w+7},
   -\epsilon\frac{3 (3 w+1) \left(9 w^2+3 w+\sqrt{9 w^2+3 w+1} \sqrt{45 w^2+51 w+5}+1\right)}{63 w^2+30 w+7},
   \right. \nonumber \\
&&\ \ \left. -\epsilon\frac{3 (w+1) (3 w+1) \left(9 w^2+12
w+1\right)}{63 w^2+30
   w+7},
    -\epsilon\frac{3 (3 w+1)
   \left(9 w^2+3 w-\sqrt{9 w^2+3 w+1} \sqrt{45 w^2+51 w+5}+1\right)}{63 w^2+30
   w+7}\right\}.\nonumber
   \end{eqnarray}
   \end{widetext}

   Finally, for the critical points $P_5^\epsilon$ the associated eigenvalues
   read
\[\left\{0,0,-\epsilon\frac{2 (3 w+1) \left(-3 w+\sqrt{3
w-2}+1\right)}{3 w-1},6\epsilon (w+1)\right\}.\]

\section{The direction $\Ex$} \label{subsE1}

In subsection \ref{localanal} we have investigated the system
\eqref{eqQ}-\eqref{eqy}, leaving outside the decoupled equation
\eqref{Aux}. However, this equation provides information of how
Kantowsky-Sachs $f(R)$ models are related to more general
anisotropic geometries. For instance from \eqref{Aux} we see that
$\Ex=0$ is an  invariant set of \eqref{eqQ}-\eqref{Aux}. This set
is closely related with the so-called ``silent boundary'' one
\cite{Matarrese:1993zf,uggla,wainwright}. \footnote{ The concept
of ``silent boundary'' was introduced for inhomogeneous
cosmologies and it is defined by the condition $e_1^1/\beta=0$
where $\beta=H+\sp$, with the $\beta$-gradient being equal to
zero. It corresponds to the divergences of $H+\sp$. In
particular, as the initial singularity is approached, the radial
part of the metric tends to zero and the null cone becomes a
timeline (see section 5.3 of
 \cite{vanElst:2001xm} for a
clear physical discussion). In our analysis we do not use
$\beta$-normalization, however, as $E_1^1=0$ is approached, a
similar physical behavior is attained (that is the radial part of
the metric tends to zero and the local null cone collapses onto
the timeline). } Note that, as current investigations suggest, in
$G_0$-cosmologies, that is in cosmologies where there is no
symmetry, both past and future attractors belong to the silent
boundary \cite{wainwright,lim}.

In the present work we do not focus in how Kantowsky-Sachs $f(R)$
are related to more general anisotropic models, and thus, we do not
elaborate completely equation \eqref{Aux}. However, we can still
obtain the corresponding significant physical information, since
the stability along the $\Ex$ direction is determined by
calculating the sign of ${\partial \left(\Ex\right) '}/{\partial
\Ex}$, which coincides with the eigenvalue associated to the
eigenvector $\Ex$. Thus, if it is negative then the small
perturbations in the $\Ex$ direction decay, while if it is
positive they enhance.

From the sign of ${\partial \left(\Ex\right) '}/{\partial
\Ex}|_{{\cal A_\epsilon}}=-\frac{\left(9 w^2+12 w+1\right)
\epsilon }{3 (3 w+1)}$ it follows that ${\cal A}_+$ (${\cal A}_-$)
is stable (unstable) to small perturbations in the $\Ex$ direction
provided $w_+<w\leq 1$, otherwise it is unstable (stable).

From the sign of ${\partial \left(\Ex\right) '}/{\partial
\Ex}|_{{\cal B_\epsilon}}= -\frac{(3 w+1) \epsilon }{3 (w-1)}$ it
follows that ${\cal B}_+$ (${\cal B}_-$) is unstable (stable) to
small perturbations in the $\Ex$ direction.

From the signs of ${\partial \left(\Ex\right) '}/{\partial
\Ex}|_{P_1^\epsilon}=2\epsilon \left(3 w+1\right)$  it follows
that $P_{1}^-$ ($P_{1}^+$) is stable (unstable) to small
perturbations in the $\Ex$ direction provided $-\frac{1}{3}< w<1,$
otherwise it is unstable (stable). Since ${\partial
\left(\Ex\right) '}/{\partial \Ex}$ vanishes at $P_2^\epsilon$ and
at $P_3^\epsilon$, it follows that they are neutrally stable to
small perturbation in the $\Ex$ direction.

By analyzing the sign of ${\partial \left(\Ex\right) '}/{\partial
\Ex}|_{P_4^\epsilon}=-\frac{3 (3 w+1) \left(9 w^2+12 w+1\right)
\epsilon }{63 w^2+30 w+7}$ we deduce that $P_4^+$ ($P_4^-$) is
stable (unstable) to perturbations in the $\Ex$ direction if
$w_+<w\leq 1$.

From the sign of ${\partial \left(\Ex\right) '}/{\partial
\Ex}|_{P_5^\epsilon}=\frac{(3 w+1) \left((3 w-1) \epsilon +2
\sqrt{3 w-2}\right)}{3 w-1},$ it follows that $P_5^+$ ($P_5^-$),
whenever exists, is always unstable (stable) to small
perturbations in the $\Ex$ direction.

Finally, note that since  \[\frac{\partial \left(\Ex\right)
'}{\partial \Ex}|_{{C_\epsilon}}=2 (n-1)  [\epsilon +\cos (u)],\]
then for $1<n\leq 3,$ $C_+$ is always unstable (except for
$u=\pi$) to small perturbations in the $\Ex$ direction, while
$C_-$ is always stable (except for $u\in \{0,2\pi\}$).

Lastly, it is interesting to mention that all the critical points
of the reduced system, except $P_{2,3}^\epsilon,$ belong  to the
``silent boundary'' in the full phase space, that is each one has
a representative in $\Psi\times\mathbb{R}$ with $\Ex=0$. In
particular, the critical points ${\cal A}_\epsilon,\, {\cal
B}_\epsilon$ correspond to isotropic silent singularities of the
full five-dimensional phase space and $P_5^\epsilon$ correspond to
an anisotropic one.

\end{document}